\documentclass[prodmode,acmtops]{acmsmall}
\usepackage{amsmath, mathrsfs, amssymb}
\usepackage[utf8]{inputenc}

\usepackage{graphicx}
\usepackage{subfloat}
\usepackage[subrefformat=parens,labelformat=parens]{subfig}
\graphicspath{ {images/} }
\usepackage{fixltx2e}
\usepackage{dblfloatfix}
\usepackage{pdflscape}
\usepackage{adjustbox}
\usepackage{rotating}

\usepackage{longtable}
\usepackage{placeins}
\usepackage{booktabs}
\usepackage[table,xcdraw,dvipsnames]{xcolor}

\usepackage{breqn}
\usepackage{multicol}

\usepackage{pslatex}
\usepackage{enumitem}

\usepackage{multirow}

\usepackage{tabularx}
\usepackage{array}
\newcolumntype{L}[1]{>{\raggedright\let\newline\\\arraybackslash\hspace{0pt}}m{#1}}
\newcolumntype{C}[1]{>{\centering\let\newline\\\arraybackslash\hspace{0pt}}m{#1}}
\newcolumntype{R}[1]{>{\raggedleft\let\newline\\\arraybackslash\hspace{0pt}}m{#1}}

\usepackage{booktabs, siunitx}
\usepackage{float}
\usepackage{url}

\DeclareGraphicsExtensions{.pdf,.png,.jpg}
\def\c#1{{\emph{#1}}}
\def\oldchange#1{#1}
\def\changeb#1{#1}

\newdef{assumption}{Assumption}

\newcommand{\SET}[1]{\ensuremath{#1}}               % Sets
\newcommand{\EV}[1]{\ensuremath{\mathscr{#1}}}                % Evidence in Bayesian
\def\PRI{{\textbf{PRI }}}

\def\P{\text{\rm Prob}}

\newcommand{\E}{\mathbb{E}}   % Expectation Value with proper E

\renewcommand{\Pr}{\mathbb{P}}         % Probability

\newcommand{\lsplitcell}[2][l]{%
  \begin{tabular}[#1]{@{}l@{}}#2\end{tabular}}

\begin{document}
% Page heads
\markboth{P Mac Aonghusa and D. Leith.}{Don't let Google know I'm lonely!}

% Title portion
\title{Don't let Google know I'm lonely}
\author{P\'{O}L MAC AONGHUSA
\affil{IBM Research and Trinity College Dublin}
DOUGLAS J. LEITH
\affil{Trinity College Dublin}}

\begin{abstract}
From buying books to finding the perfect partner, we share our most intimate wants and needs with our favourite online systems. But how far should we accept promises of privacy in the face of personalised profiling?  In particular, we ask how can we improve detection of sensitive topic profiling by online systems? We propose a definition of privacy disclosure we call $\epsilon$-indistinguishability from which we construct scalable, practical tools to assess learning potential from personalised content. We demonstrate our results using openly available resources, detecting a learning rate in excess of $98\%$ for a range of sensitive topics during our experiments.
\end{abstract}


\terms{Privacy, Interference, Recommender-System, Search}


\acmformat{P\`{o}l Mac Aonghusa and Douglas J. Leith, 2015. Don't let Google know I'm lonely!}

\begin{bottomstuff}
Supported by Science Foundation Ireland grant 11/PI/11771.

Author's addresses: P. Mac Aonghusa, IBM Research, IBM Technology Campus, Dublin, Ireland; Douglas J. Leith, School of Computer Science and Statistics, Trinity College Dublin, Ireland.
\end{bottomstuff}

% Copyright
\setcopyright{acmcopyright}
\issn{1094-9224/2015} 
\doi{0000001.0000001}

%% Metadata Information
\acmArticle{X}
\acmVolume{0}
\acmNumber{0}
\acmYear{2016} 
\acmMonth{05} 

\maketitle

%%%%%%%%%%%%%%%%%%%%%%%%%%%%%%%%%%%%%%%%%%%%%
%%%%%%%%%%%%%%%%%%%%%%%%%%%%%%%%%%%%%%%%%%%%%

%%%%%%%%%%%%%%%%%%%%%%%%%%%%%%%%%%%%%%%%%%%%%
\begingroup
\let\clearpage\relax                 % Suppress pagebreak

\section{Introduction}% 
\label{sec:intro}
We investigate threats to user privacy due to \emph{inference} by a search engine through users' online behaviour. Of particular interest is learning related to potentially sensitive topics such as health, finance and sexual orientation. Our goal is to inform the user by \emph{detecting} evidence of privacy disclosure through analysis of personalised content. Our method is readily implementable with available open tools, simple to apply, and provides highly accurate results.

Our approach is borrowed from \emph{black-box} testing: given a sequence of user queries we embed a subsequence of \emph{probe queries} and observe corresponding search engine responses. By analysing changes in the responses to the probe queries over time, we hope to be able to spot learning of topics the user considers sensitive and so would prefer not to disclose to the search engine. 

Using Bing and Google Search, we demonstrate that by monitoring changes in the adverts displayed in the response to probe queries we are able to accurately detect evidence of learning for a range of sensitive topics in over $98\%$ of cases.  Topics studied include medical conditions (cancer, anorexia \emph{etc}), sexual orientation, disability, bankruptcy and unemployment.  Our method is accurate, with typical false detection rates of less than $10\%$ (and less than $1\%$ for many sensitive topics). We also show that detection rates remain high for anonymous users, suggesting that search engines learn quickly; even without search history as background knowledge. Our estimation of search engine adaptation rates indicate that sensitive topic learning is detectable after as few as $3-4$ queries on average.

The main contributions in this paper are:
\begin{list}{\quad }{}
    \item A definition of privacy we call \oldchange{$\epsilon$-indistinguishability} that is both compatible with existing privacy models and readily implementable as a practical user technology
    \item An effective method for change detection across a sequence of queries by collecting and comparing  responses to a subsequence of preselected \emph{probe queries}
    \item A fast, scalable estimator of \changeb{$\epsilon$-indistinguishability}, we call \PRI \changeb{(''\textbf{PR}ivacy for \textbf{I}ndividuals'')} and \oldchange{which we implement using standard tools and apply in subsequent experiments}
    \item An extensive measurement campaign showing that evidence of adaptation is easy to detect for a wide range of sensitive topics.
\end{list}  

\changeb{In this paper we focus on raising awareness of privacy concerns arising from personalisation by online services. We position our contribution as a starting point in a multi-step program. Our goal here is to demonstrate that practical individual awareness is possible, providing a stepping stone toward effective counter-measures.}

\section{Related Work}

Personalisation of web search through implicit data collection -- for example location, IP address and browser agent -- is well studied. See \cite{decade:web:mining} for an historical survey of results in web mining for personalisation. Even so-called `private' browsing mode may not suffice; in \cite{Aggarwal:2010:APB:1929820.1929828}, the authors investigate how a range of popular browser extensions and plugins undermine the security of private browsing. 

\changeb{In \cite{boyd2011talk}, individual user concerns with privacy are viewed in terms of two major factors -- awareness of a sensitive social situation, and, the ability of an individual to control the social situation. In this paper we focus on raising awareness through detection of profiling of sensitive topics. Effective counter measures allowing a user to control their exposure to profiling are outside the scope of this present paper.}
\changeb{
Evidence that users are sensitive to personalisation -- and will respond to increased awareness -- is given in \cite{Panjwani:2013:UPD:2470654.2466470}, where a user study of internet search users showed a slight preference for personalised content. After raising awareness \emph{by fully informing} users about risks to their privacy, the majority of users were satisfied to forego personalisation when search topics were judged sensitive. In \cite{Agarwal:2013:ERU:2501604.2501612}, a larger user study explores privacy concerns in more depth, finding that users are more concerned about the potential of being shown suggestive or embarrassing content than they are of tracking.} 
  
While improved user experience is generally offered as a positive motivation for user profiling, for example, see the Google online privacy policy \cite{google:policy}, negative associations have been reported in the research literature:
\begin{description}
\oldchange{
\item [Discrimination] Negative consequences associated with personalisation are investigated in \cite{Sweeney:2013:DOA:2460276.2460278}, where an extensive review of adverts from Google and \url{Reuters.com} showed a strong correlation between adverts suggestive of an arrest record and an individual's ethnicity. Searches containing first names considered black-identifying were on average $25\%$ more likely to receive adverts indicative of an arrest record than searches including white-identifying first names. In \cite{Guha:2010:CMO:1879141.1879152}, the authors identified more than half of online advertising targeted \emph{exclusively} to gay men was neutral to sexual-orientation -- posing a privacy threat through inadvertant user clicking.
Our experiments confirm that evidence of profiling that could be termed discriminatory is detectable with high confidence. 
}

\oldchange{
\item [Censorship] Personalisation as a form of censorship --  termed a {filter bubble} in \cite{Pariser:2011:FBI:2029079} -- is explored in \cite{Hannak:2013:MPW:2488388.2488435}. In a filter bubble, a user cannot access subsets of information because the recommender system algorithm has decided it is irrelevant for that user. In \cite{Hannak:2013:MPW:2488388.2488435} a filter bubble effect was detected in the case of Google Web Search in a test using 200 users. Censorship is an interesting complimentary perspective on user profiling that is not considered in this paper. 
%Our experiments were run from multiple computers located in the same University subnet where differentiated localisation effects such as those reported in \cite{Hannak:2013:MPW:2488388.2488435} are less likely to occur. An interesting topic for future research would be to apply the methods used in \cite{Hannak:2013:MPW:2488388.2488435} with the tools developed in the present paper to quantify the filter bubble effect further.
}
\end{description}

Much of the research literature in recommender system privacy has focused on techniques to \emph{implement privacy}; for data processing on the server side -- or -- on the user side, techniques to obfuscate, encrypt or otherwise hide queries or mask user identity from the recommender. 
The query masking or obfuscation technique has been explored extensively, see for example \cite{howe2009trackmenot,peddinti2011limitations}. The essential challenge in this type of approach is to define a practical method of selecting the `noise' queries \oldchange{and clicks} to provide a verifiable level of anonymity while not overly upsetting overall utility. In our experiments  we observe the recommender systems responding quickly to changes in user topic, compensating quickly after topic changes so that `noise injection' alone may not provide adequate protection. 
%TrackMeNot, for example, has extended their original implementation based on `noise injection' to include simulated clicking as an additional source of obfuscation.

Several authors have examined privacy in the context of recommender systems, broadly addressing the question of data privacy with respect to the user data once collected by the recommender. In a 2001 work, \cite{rama01pr}, the authors identify threats from data linking or combination by identifying similar patterns of preference or behaviour conjunction with other data sources to uncover identities and reveal personal details. In their concluding remarks, the authors state that ``the ideal deterrents are better awareness of the issues and more openness in how systems operate in the marketplace. In particular, individual sites should clearly state the policies and methodologies they employ with recommender systems''.

\oldchange{
Techniques for preserving individual user privacy or anonymity have also been extensively discussed in the literature. A typical approach is to apply encryption and multi-party computation techniques to process sensitive user queries, leveraging techniques from the privacy preserving data mining domain. For example, in \cite{IEEE:so87258,erkin2011ef}, the authors propose to encrypt privacy sensitive data and generate recommendations by processing them under encryption. Approaches of this type typically rely on a user, or a learning algorithm,  being able to identify {which} queries are sensitive, and trust in the service provider to perform query processing under secure encryption. In contrast, our approach seeks to inform a user about their ongoing privacy status. }
%During the training phase a user is asked to label query sessions as sensitive or non-sensitive rather than individual queries. Session level labeling was adopted in our experiments for two reasons; a) individual query labeling places an unreasonable burden on the user, and, b) query session level labeling seems a better representation of overall user intent, allowing for small variations when a user chooses individual queries.

\oldchange{
Accountability, and enforcement of accountability, for privacy policy is an active area of research. Regulatory requirements for data handling in industries such as Healthcare (HIPPA) and Finance (GLBA) are well established. The position with respect to handling of data collected by online recommender systems is less clear. In \cite{Datta:2014:PTA:2769818.2769824}, the author reviews computational approaches to specification and enforcement of privacy policies at large scale. Our approach differs in design, being intended for individual user implementation. While we also make use of inference, it is statistical inference rather than the logic-based inference approach discussed in \cite{Datta:2014:PTA:2769818.2769824}. 
}

\oldchange{
Tools that seek to inform a user about whether an adversary is potentially gathering and sharing information exist. Browser add-ons such as Mozilla Lightbeam, \cite{lightbeam}, and PrivacyBadger, \cite{badger}, show a user where data is shared with third parties from sites they visit. XRay, \cite{Lecuyer:2014:XEW:2671225.2671229}, reports high accuracy in identifying which sources of user data such as email or web search history might have {triggered} particular results from online services such as adverts. }

\changeb{
Privad, \cite{Guha:2011:PPP:1972457.1972475}, and ObliviAd, \cite{Backes:2012:OPS:2310656.2310685} allow advertisers to provide private advertising, by employing intermediate agents to mask direct access to user information. In both cases external agents hide user intentions and interactions with adverts. ObliviAd use secure hardware-based private information retrieval for advert distribution, with advertiser billing provided by secure tokens. These features allow the authors to provide provable cryptographic guarantees of anonymity for users of the system. 
}

%Our approach applies several similar ideas to XRay such as differential comparison of outputs. While  XRay attempts to explain potential sources for online personalisation, we attempt to inform a user when learning of potentially sensitive topics by a recommender system appears to have occured.  Our implementation -- we call \PRI -- is also considerably different to XRay. XRay is an online service called from a user's computer, whereas we implement entirely within a single user's computer. 

\oldchange{
The importance of adverts to commercial search engines is underlined by the volume of research into how to position adverts to direct user attention to them.
The position of commercial content on a result page is a key determinant of whether an \oldchange{advert} will be selected or not. For example, in \cite{jansen2013effect}, the authors determined that the first \oldchange{advert} on a page accounts for approximately $80\%$ of commercial revenue. In \cite{richardson2007predicting}, the authors include results from user eye-tracking demonstrating how dramatically a user focuses on the first few results on a page.
}

\oldchange{
Recommender systems continue to evolve more sophisticated methods of content selection. Semantic targetting techniques, where the overall theme of a web-page is used to select contextually related adverts, are used by companies such as Google Knowledge Graph, \cite{google:graph}, and iSense, \cite{isense}, while Zemanta, \cite{zemanta}, provides a browser add-on that suggests semantically relevant articles, images, links and keywords to content creators. Experimental results, with Google in particular, appear to show early promise that the \PRI technique applies to content selection models other than keyword-based .     
}

\oldchange{
Modelling a system as a black-box, where internal details of recommender systems algorithms and settings are unknown to users, is mentioned in several sources \cite{Datta:2014:PTA:2769818.2769824} and \cite{Hannak:2013:MPW:2488388.2488435}. While the term ``black-box'' is used extensively throughout the privacy literature, the present authors have not seen black-box testing techniques -- such as employing probes to calibrate responses -- using extensively in the privacy literature.
}

In this paper, we seek a practical approach, allowing a user {inform} themselves of privacy disclosure threats due to search engine inference.  By exploiting the idea that a commercial search engine  combines user data with background knowledge to create personalised recommendations we ask how to assess ongoing risks to privacy based on simple observation of the recommender?  In this way we help a user to make reasonable assessments of the ongoing risk of disclosure while using a search engine.

\section{Threat Model}
\label{sec:profit}
We consider privacy for a class of commercial internet systems seeking to maximise expected revenue by personalising commercial content to attract user interest.  Users are generally aware of good privacy practice for obviously personally identifying information -- such as name, address and credit card numbers -- reflecting the visibility of personal identifiability as a central concept in information privacy regulation \cite{schwartz2011pii,ohm2010broken}. Personalisation practices, based on obvious features such as location, IP address and browser identifier, are also well known and extensively discussed in the literature \cite{decade:web:mining}. \oldchange{In this paper we focus on detecting evidence of recommender system learning.  User traceability or pseudo-anonymity is not considered.}

It seems reasonable to assume that a for-profit commercial search engine selects page content to maximise its expected revenue. This means that when a search engine infers that a particular advertising \oldchange{topic} is likely to be of interest to a user, and so more likely to generate click through and sales, it is obliged to use this information when selecting which adverts to display.  

In this context the threat model we consider is one of \emph{distinguishability} rather than individual identifiability -- a search engine does not seek to identify  \oldchange{the user as an individual} but rather it seeks to determine \oldchange{the user's likely interest in} commercially valuable \oldchange{topics}.
Privacy becomes an issue when any of the \oldchange{topics} matches \oldchange{subjects} deemed sensitive by the user. 

Since a revenue maximising search engine acts to display adverts associated with topics \oldchange{it detects} are most interesting to the user, the potential exists to detect search engine learning via analysis of changes in the choice of displayed adverts and to inform the user of this learning.
 
In our experiments we find that adverts do indeed provide sufficiently dynamic content, as we shall show  \oldchange{in Section~\ref{sec:feature:select}}.  We note that, \oldchange{over the period of short query sessions we consider here,} page content is usually constrained to be relatively insensitive with respect to personalisation in order to provide so-called rank-stability; link-based search algorithms are termed rank-stable if small perturbations in the link structure of the input graph do not affect the output ranking order it produces, \cite{lempel2005rank} and \cite{Langville:2006:GPB:1146372}.  In contrast, adverts may be chosen relatively freely, a fact which has also been noted by other authors, for example see \cite{Guha:2010:CMO:1879141.1879152}. 

\oldchange{Evidence that page content varies for individuals over periods of time has been reported. \cite{Hannak:2013:MPW:2488388.2488435} also observes that localised content such as news items is necessarily volatile and will vary to reflect currency. Our focus, in this paper, is on detecting learning by observing adverts and we do not consider other personalised content such as news and weather specifically in this study. For the typical period of time of a typical user session in this study -- typically less than 20 minutes -- experimental results in Section~\ref{sec:feature:select} indicate the assumption of ``short-term'' rank-stability in page content considered in our study is reasonable.
}

%%%%%%%%%%%%%%%%%%%%%%%%%%%%%%%%%%%%%%%%%%%%%
\section{Mathematical Formulation}

%%%%%%%%%%%%%%%%%%%%%%%%%%%%%%%%%%%%%%%%%%%%%
\subsection{$\epsilon$-indistinguishability}
We assume that a user interacts with a search engine by issuing a query, receiving a web page in response and then clicking on one or more items in the response. A single such interaction, labeled with index $j$, consists of a \emph{query, response page, item-click} triple, denoted \oldchange{$\Omega_j = \left(q_j, p_j, l_j \right)$}.
A user session of length $k>0$ steps consists of a {sequence} of $k$ individual \oldchange{steps,} and is denoted  \oldchange{$\left\{\Omega_k\right\}_{k\ge1}$}.   
The sequence of interactions \oldchange{$\left\{\Omega_k\right\}_{k\ge1}$} is jointly observed by  the user and the search engine -- and perhaps several other {third-party observers}. 

Let $\EV{E}_k$ denote the {prior evidence} -- also referred to as {background knowledge} -- available to an observer at the start of step $k$.  We assume the search engine does not change its background knowledge during a user session other than through $\Omega_k$.   That is, $\EV{E}_1$ denotes the prior evidence available to an observer before the user session begins, \emph{e.g.} the user's login profile, historical queries, weblogs  \emph{etc}, and for $k=2,3,\cdots$ we have,
\begin{align}
\EV{E}_k &= \{\EV{E}_{k-1}, \Omega_{k-1} \} \notag
\end{align}%
%\oldchange{
%\noindent It is impractical for a user to determine the exact form of $\EV{E}_1$ when they do not have access to the internal workings of the search engine. How to model $\EV{E}_1$ when implementing \PRI is the subject of Assumption \ref{a:1}.
%}

Let $\SET{C} = \{c_1, \ldots, c_N\}$ denote a set of interest categories which the {user considers to be sensitive}, \emph{e.g.} \c{bankrupt}, \c{cancer}, \c{addiction}, \emph{etc} and gather all non-sensitive interest categories into a catch-all category denoted $\bar{c}$.   The category in which the user is interested in the current session is a random variable $X_{c}$ taking values in $\SET{C}\cup\{\bar{c}\}$. 
\oldchange{In subsequent experiments multiple sensitive user categories will sometimes be aggregated into a single sensitive topic for clarity, so that $\SET{C}\cup\{\bar{c}\} = $\{\c{sensitive}, \c{non-sensitive}\} in this simplified case.}
\oldchange{
The term ``category''  here denotes a topic or theme of interest to the user. Since the search engine adversary is regarded as a black-box we do not know if such {user categories} correspond to internal classifications by the adversary.  Our intent is to detect if there is evidence of learning about {user categories} of interest to the user. We do not attempt to understand what internal classifications -- if any -- the search engine might apply to the user. 
}

We adopt an {indistinguishability} definition of {disclosure risk}, tailored to our particular context:
\begin{definition}[ $\epsilon$-indistinguishability ]
A user session $\Omega_k$ satisfies  \oldchange{$\epsilon$-indistinguishability}  with respect to sensitive category $c \in \SET{C}$ \oldchange{if there exists an $\epsilon > 0$ such that}
\begin{align}
M_k(c) \le e^\epsilon, \ \ k=1,2,\cdots
\end{align}
%\normalsize
where 
\begin{align}
M_k(c) := \frac{\P(X_c = c | \Omega_k, \EV{E}_{k} )}{\P(X_c = c | \EV{E}_{1} )}
\label{eqn:epsilon:identify}
\end{align}
In other words there is \emph{a posteriori} indistinguishability of interest in sensitive category $c$ after observing $\Omega_k$, $k=1,2,\cdots$.  
\label{def:indistinguish}
\end{definition}

Given the sequence of observations \oldchange{$\{\Omega_k\}_{k\ge1}$} our aim is, with high probability, to (1) determine whether \oldchange{$\epsilon$-indistinguishability} has been violated for one or more of the sensitive categories in $\SET{C}$ (and so the adversary is likely to have successfully learned about the users interest in one or more of these categories), and (2) identify which of these sensitive categories have been learned.

%%%%%%%%%%%%%%%%%%%%%%%%%%%%%%%%%%%%%%%%%%%%%
\subsection{Using Probe Queries to Simplify Estimation}
\label{sec:priv:disclose:search}
Estimating $M_k(c)$ is challenging since it depends on the full user session history \oldchange{$\{\Omega_j\}_{j=1,\dots,k}$} up to step $k$.   To simplify the task we assume that the user issues a pre-defined {probe} query at intervals during the session and that the links in the response to this query are not clicked.  
\oldchange{
In brief, a probe query should be \emph{plausible} in relation to a sensitive topic so that it does not suggest a change of topic to the search engine; a probe query should  also be \emph{ambiguous} so that the search engine has several possible adaptations to the probe query. In Section~\ref{sec:probe:queries} experimental probe query selection is discussed, where selecting high-frequency terms appearing on multiple search result pages, while taking care to avoid obviously revealing terms, is shown to be a practical method of probe selection.
}
In practice, a probe query might be issued in an automated manner by the user's browser and the response processed in the background so as not to disturb the user.  

In addition, we make the following assumptions.

\begin{assumption}[Informative Probe Query]\label{a:1} 
Let $\SET{K}\subset\{1,2,\cdots\}$ label the subsequence of steps at which a probe query is issued.   
At each step $k\in\SET{K}$ at which a probe query is issued,
\begin{align}
\frac{\P(X_c = c | \Omega_k, \EV{E}_{k} )}{\P(X_c = c |  \EV{E}_{1} )}&=\frac{\P(X_c = c | \Omega_k,\EV{E}_{1} )}{\P(X_c = c|  \EV{E}_{1})}
\label{eqn:assumption:informative}
\end{align}
\end{assumption}
\noindent \oldchange{That is, it is not necessary to explicitly use knowledge of the search history during the current session when estimating $M_k(c)$ as this is already reflected in the response to the probe query at step $k$.}  Assumption \ref{a:1} greatly simplifies estimation as it means we do not have to take account of the full search history, but requires that the response to the probe query reveals any search engine learning of interest in sensitive category $c$ which has occurred. Methods for the selection of an appropriate probe query that tends to elicit revealing responses are discussed in detail \oldchange{in Section~\ref{sec:probe:queries}}. 

%\oldchange{
%The practical impact of Assumption \ref{a:1} is in simplifying the task of choosing a training set during implementation. A second benefit of Assumption \ref{a:1} is in overall scalability of \PRI. As time elapses it is important to refresh the training data with recently observed adverts. For example, comparison of two samples of data gathered for Google six months apart, with identical setup and query scripts, reveals that $36\%$ of adverts from the older data set continue to appear in the newer data set. While the same comparison for for Bing the surviving advert content after six months is $17\%$.
%Assumption \ref{a:1} is important for implementation, stating that `older' adverts in the training data may be replaced with new adverts, helping to contain memory growth of over time. We will retain $\EV{E}_1$ explicitly in subsequent expressions to emphasise that we must make an explicit choice of an initial training set at the implementation stage.}

\begin{assumption}[Revealing Adverts]\label{a:2} In the search engine response to the probe query at step $k$ it is the adverts ${a}_k$ on response page $p_k$ which primarily reveal learning of sensitive categories.   Therefore, since the probe query is fixed and the response page -- for probe queries -- is not clicked by the user,
\begin{align}
\P(X_c = c | \Omega_k,\EV{E}_{1}) = \P(X_c = c | {a}_k,\EV{E}_{1} ), \ k\in \SET{K}\label{eq:assum0}
\end{align}
\end{assumption}
\noindent  \oldchange{This is in line with the observation in Section~\ref{sec:profit} that, over the lifetime of a typical user session, the informational response (search result links etc) to a query tends to be insensitive to learning of user interests.   Note that personalisation of search results over longer time scales e.g. in response to changes in geographic location may still occur.}

%\oldchange{In Section~\ref{sec:time:learn}, we report a `lag effect' where adaptation to a new topic appears to be slightly lower over the first $1-2$ probe queries at the beginning of a query session when the topic has changed. The experimental results in Section~\ref{sec:time:learn} show that the lag effect disipates over the life of multi-probe query session. Assumption \ref{a:2} holds, therefore, in a general sense, allowing for slight, initial variation while the recommender system appears to confirm or refine its interpretation of a topic.}

Under Assumptions \ref{a:1}-\ref{a:2}, when
\begin{align}
{M}_k(c)=\frac{\P(X_c = c | {a}_k,\EV{E}_{1} )}{\P(X_c = c|\EV{E}_{1})} > e^\epsilon
\end{align}
for \emph{any} $k\in \SET{K}$ then \oldchange{$\epsilon$-indistinguishability} is violated.   To ensure that the converse holds, namely that when ${M}_k(c) \le e^\epsilon$ for all $k\in \SET{K}$ then \oldchange{$\epsilon$-indistinguishability} is satisfied, we also need the following assumption.

\begin{assumption}[Sufficiency of Sampling]\label{a:3}
When $M_k(c) \le e^\epsilon$ for $k\in \SET{K}$ then $M_k(c) \le e^\epsilon$ for $k\in\{1,2,\cdots\}$.  That is, when  \oldchange{$\epsilon$-indistinguishability} is satisfied at the subsequence of steps $\SET{K}$ at which the probe query is issued then it is satisfied at all steps. 
\end{assumption}
\noindent  In practice it can be difficult to verify whether Assumption \ref{a:3} holds or not.  When we cannot rely on Assumption \ref{a:3} then, as already noted, violations ${M}_k(c)> e^\epsilon$ for $k\in\SET{K}$ are still informative of disclosure risk and analysis based on the values ${M}_k(c)$, $k\in \SET{K}$ should be regarded as an underestimate, or lower bound, of disclosure risk for the user.
\subsection{Advert Text Processing}
\oldchange{
We briefly summarise Natural Language Processing (NLP) techniques we use to preprocess advert text into a format for analysis. For a full treatment of these techniques, using the Python language, see \cite{Bird:2009:NLP:1717171}. 
}

\oldchange{In the first preprocessing step, advert text from result pages is extracted and tokenised into individual words by using white-spaces and punctuation as token separators. Common, uninformative high-frequency {stop-words} are removed  and {stemming} is performed to remove common prefixes and suffixes, for example, $\{\text{clicking, clicks, clicked}\} \rightarrow \{\text{click}\}$ . 
}
\oldchange{After preprocessing and during the training phase we assign each unique keyword an integer id and also count the keyword's frequency of occurrence in the training set to produce a lookup table of keyword id-frequency pairs. The lookup table of keywords and associated occurence counts is generally termed a {Term-Frequency} (TF) representation in NLP. Because the previous preprocessing step has removed uninformative high frequency words and has reduced word variations through stemming, the TF table transforms advert text to a smaller more computable representation.
}

\oldchange{During subsequent testing, an advert encountered on a result page is first preprocessed and then mapped to keyword ids the TF table. Preprocessing an advert in this way is called {vectorisation} in NLP, where the $\text{n}^{th}$ component of the vectorised advert is the number of occurences in the advert of the keyword stem with integer id $\text{n}$. Given two vectorised adverts it is now possible, for example, to compute similarity measures between adverts using any convenient vector metric -- such as Cosine Similarity which is effectively vector inner product.
}

To streamline notation we assume that the available advert locations are the same for all pages $p_k$, $k=1,2,\cdots$.  Let ${a}_k=\{a_{1,k},\cdots,a_{n,k}\}$ be the set of adverts on page $p_k$.  An advert $a_{i,k}=\{w_1, w_2,\cdots, w_{|a_{i,k}|}\} $ \oldchange{appearing on a result page is modelled as} a sequence of words drawn from \oldchange{a set of }natural language \oldchange{terms} $\SET{D}$ \oldchange{(as discussed later, in our work we derive this set from training data)}.  In practice result pages may contain \oldchange{localised content, such as news and weather, as well as other format content such as } images and links, but leave consideration of these to future work.

We represent text preprocessing operations such as stemming and stop word removal operations  as a filter over $\SET{D}$. Specifically, let $f:\SET{D}\rightarrow\tilde{\SET{D}}\cup\emptyset$ be a text processing/filtering map taking words in dictionary $\SET{D}$ either to a content-bearing \emph{term} in set $\tilde{\SET{D}}$  or to a null term $\emptyset$ corresponding to deletion of the word.   
Let  $\tilde{a}_{i,k}=\{f(w_1), f(w_2), \ldots, f(w_{|a_{i,k}|} ) \}$ denote the sequence of content-bearing terms obtained by applying map $f$ to advert $a_{i,k}$ and let ${\tilde{a}}_k=\{\tilde{a}_{1,k},\cdots,\tilde{a}_{n,k}\}$. 

Filtering implies a subtle assumption common in text classification, namely that  preprocessing and filtering only removes statistically uniformative content that is not significant for subsequent analysis.  That is, for any advert ${a}_k$ and category $c$, 
\begin{align}
\P(X_c= c |{a}_k )&=\P(X_c=c |\ {\tilde{a}}_k)
\end{align}%
In addition, we assume that the content-bearing terms in $\tilde{\SET{D}}$ have distinct meanings in the following sense. 

Let $\Delta_{w,k}$ be a random variable which takes value $1$ when a term drawn uniformly at random\footnote{
Precisely, let $b_k$ denote the sequence of length $\sum_{i=1}^n|\tilde{a}_{i,k}|$ obtained by concatenating sequences $\tilde{a}_{i,k}$, $i=1,\cdots,n$.  Select an index uniformly at random from $\{1,\cdots,\sum_{i=1}^n|\tilde{a}_{i,k}|\}$ and if the term in $b_k$ at that index equals $w$ then $\Delta_{w,k}=1$, else $\Delta_{w,k}=0$.} 
from the page adverts ${\tilde{a}}_{k}$ equals $w$ and $\Delta_{w,k}$ equals $0$ otherwise.  Hence, when $\Delta_{w,k}=1$ then term $w$ is known to be contained in the adverts displayed on a page.
Our assumption is that after text processing $\P( X_c = c | {\tilde{a}}_k, \EV{E}_{1})$ possesses the following mixture form:
\begin{assumption}[Text Processing]\label{a:4}
\begin{align}
%\P( X_c = c |\Delta_{w,k}=1, \tilde{{a}}_{k}, \EV{E}_{1} )=\P( X_c = c |\Delta_{w,k}=1, \EV{E}_{1} )
&\P( X_c = c | {\tilde{a}}_k, \EV{E}_{1})%\notag \\
%&= \sum_{w\in\tilde{\SET{D}}} \P( X_c = c |\Delta_{w,k}=1,{a}_k,  \EV{E}_{1})  \P(\Delta_{w,k}=1|{a}_k,  \EV{E}_{1}) \notag \\
%&\quad
= \sum_{w\in\tilde{\SET{D}}} \P( X_c = c |\Delta_{w,k}=1,  \EV{E}_{1}) \P(\Delta_{w,k}=1|{\tilde{a}}_k,  \EV{E}_{1}) \label{eq:assum2}
\end{align}
\end{assumption}
%which essentially states that each content-bearing term $w$ contributes to a distinct aspect of category $c$.   
%
\oldchange{Assumption \ref{a:4} states that categories and adverts are correlated through the keyword terms. It is a similar to the assumption underpinning techniques such as Latent Class Analysis and Collaborative Filtering in recommender systems \cite{Ricci:2010:RSH:1941884}.}

To streamline notation, from now on we omit the tilde $\,\tilde{}\,$ denoting filtered text, noting that all text is assumed to be filtered through the mapping $f$ unless stated otherwise.

%%%%%%%%%%%%%%%%%%%%%%%%%%%%%%%%%%%%%%%%%%%%%
\subsection{Bayesian Estimator}
\label{sec:bayesian:formulation} 

By Assumptions \ref{a:1}, \ref{a:2} and \ref{a:4} we have:
\begin{align}
{M}_k(c)&= \sum_{w\in{\SET{D}}}\frac{ \P( X_c = c | \Delta_{w,k}=1 ,  \EV{E}_{1}) \P(\Delta_{w,k}=1 | {a}_k ,  \EV{E}_{1})}{\P( X_c = c  |  \EV{E}_{1})}\notag &\\
&\stackrel{(a)}{=} \sum_{w\in{\SET{D}}}\frac{\P(\Delta_{w,k}=1  |   c ,  \EV{E}_{1})\P(\Delta_{w,k}=1|{a}_k ,  \EV{E}_{1}) }{\P(\Delta_{w,k}=1 |   \EV{E}_{1})}\\
&= \sum_{w\in{\SET{D}}}\frac{p_{c ,  \EV{E}_{1}}p_{{a}_k ,  \EV{E}_{1}} }{p_{\EV{E}_{1}}}
\label{eqn:m_k_c:step2}
\end{align}
where equality $(a)$ follows from Bayes Theorem and 
\begin{align}
p_{ c ,  \EV{E}_{1}}&:=\P(\Delta_{w,k}=1  |   c ,  \EV{E}_{1})\\
p_{{a}_k ,  \EV{E}_{1}}&:=\P(\Delta_{w,k}=1|{a}_k ,  \EV{E}_{1})\\
p_{ \EV{E}_{1}}&:=\P(\Delta_{w,k}=1 |   \EV{E}_{1})
\end{align}
%
%\textbf{** Estimator notation!!}
%We shall use the notation $\hat{X}$ to indicate that $\hat{X}$ is a statistical estimator for a quantity $X$. For complicated expressions we will use the notation $X \hat{=} Y$ to denote that $X$ is estimated by $Y$ -- i.e. when $Y=\hat{X}$.

We can define empirical estimators for $p_{ c ,  \EV{E}_{1}}$, $p_{{a}_k ,  \EV{E}_{1}}$ and $p_{ \EV{E}_{1}}$ in the obvious way, as follows.   
Assume the availability of a training data set $\SET{T}$ consisting of (label,advert) pairs, where the label is the category in $\SET{C}\cup\{\bar{c}\}$ with which the corresponding advert is associated.  Approximate the prior evidence at the begining of the query session empirically with this training data: $\hat{\EV{E}}_1 = \SET{T}$. 
Text preprocessing of $\SET{T}$ by filtering produces a dictionary $\SET{D}$. 
Given the dictionary $\SET{D}$, let $\phi_{\SET{D}}(x | X)$ denote the frequency with which an item $x \in \SET{D}$ occurs in sequence $X=\{x_1,x_2,\cdots,x_{|X|}\}$. That is,
\begin{align}
\phi_{\SET{D}}(x  | X) &= 
\begin{cases}
\frac{\vert \{i:i \in \{1,\cdots,|X|\},x_i = x\} \vert}{\vert X \vert}
& \mbox{if } x \in \SET{D}  \\
0 
&\mbox{if } x \notin \SET{D}
\end{cases} 
\label{eqn:phi:define}
\end{align}
%
%Adverts consist of short pieces of text of similar length. It is convenient to create a larger sample for analysis by grouping a vector of adverts from a result page into a single sequence of terms by concatentating each component advert together. Specifically, given a vector of adverts ${a}_k$, 
%let ${\CAT{{a}_k}}$ be the sequence of length $\sum_{i=1}^n|a_{i,k}|$ obtained by concatenating individual component sequences $a_{i,k}$, $i=1,\cdots,n$.
%
From the definition of $\Delta_{w,k}$ we can now define the following estimator $\hat{p}_{{a}_k , \hat{\EV{E}_{1}}}$ for $p_{{a}_k , \hat{\EV{E}_{1}}}$: 
\begin{align}
\hat{p}_{{a}_k , \hat{\EV{E}_{1}}} %&= \P(\Delta_{w,k}=1|\ {a}_k , \SET{T}) \label{eqn:phi:w:advert:step1} \\
&= \sum_{{a}\in {a}_k} \phi_{\SET{D}}(w | {a})
%\\
%& \hat{=} \phi_{\SET{D}}(w | \CAT{{a}_k})
\label{eqn:phi:w:advert}
\end{align}
%\noindent which can be readily evaluated.
%
% and estimation \eqref{eqn:phi:w:advert} by taking the concatenation of adverts on a page -- where we assume adverts are of similar length.
%
%\textbf{** We can sum over just $\phi_{\SET{D}}(w | {a})$ on \eqref{eqn:Pr_w_ak:step2} above -- because $\phi$ includes the dependency on the terms belonging to the training set -- therefore we don;t need to have $\phi(a | T)$ any more.}
%
%Observe that 
%\begin{align}
%\P(\Delta_{w,k}=1 | {a}_k , \ \EV{E}_{1})
%&= \sum_{{a}\in \SET{A}} \P(\Delta_{w,k}=1|{a} , \EV{E}_{1})\P({a})\\
%&= \sum_{{a}\in \SET{A}} \phi_{\SET{D}}(w   |  {a})\P({a})
%\end{align} 
%where $\SET{A}$ is the set of available adverts.
%
%An empirical estimator for $\P(\Delta_{w,k}=1 | {a}_k ,  \EV{E}_{1})$ is then $\sum_{{a}\in \SET{D}} \phi_{\SET{D}}(w | {a})\phi_{\SET{D}}({a}|\SET{T})$.
Similarly, we define the following empirical estimator $\hat{p}_{\hat{\EV{E}_{1}}} $ for $p_{\hat{\EV{E}_{1}}}$:
\begin{align}
\hat{p}_{\hat{\EV{E}_{1}}} &= \sum_{a\in \SET{T}} \phi_{\SET{D}}(w | {a})
\end{align}
\noindent where the sum is over adverts in $\SET{T}$. 
Letting $\SET{T}(c)$ denote the subset of $\SET{T}$ where the label is category $c$, we also obtain an empirical estimator $\hat{p}_{c , \hat{\EV{E}_{1}}}$ for $p_{c , \hat{\EV{E}_{1}}}$:
\begin{align}
 \hat{p}_{c , \hat{\EV{E}_{1}}}&= \sum_{a\in \SET{T(c)}} \phi_{\SET{D}}(w | {a})
%\sum_{{a}\in \SET{T}(c)} \phi_{\SET{D}}(w | {a}) %\phi_{\SET{D}}({a}|\SET{T}(c))
\end{align}

Combining these estimates using (\ref{eqn:m_k_c:step2}) then yields the following estimator for $M_k(c)$:
\begin{align}
\hat{M}_k(c) &= \sum_{w\in{\SET{D}}}\left( \frac{\sum_{a\in \SET{T}(c)} \phi_{\SET{D}}(w | {a})}{ \sum_{a\in \SET{T}} \phi_{\SET{D}}(w | {a}) } \cdot \sum_{{a}\in {a}_k} \phi_{\SET{D}}(w | {a}) \right) \label{eqn:mk001}
\end{align}

We refer to the expression for $\hat{M}_k(c)$ as the \textbf{PRI} estimator.

%%%%%%%%%%%%%%%%%%%%%%%%%%%%%%%%%%%%%%%%%%%%%
\subsection{Example}
\label{sec:math:discussion}
%To apply \eqref{eqn:estimate:mk:simple} in practice, consider the following model.  We suppose a dictionary $\SET{D}$ consisting of $M$ statistically significant terms $\{w_1, \ldots, w_M\}$ is provided.  
%
%Given a training set $\SET{T}$ let 
%\begin{align*}
%t_i &=\sum_{{a}\in \SET{T}} \phi_{\SET{D}}(w_i | {a}), \ i=1, \ldots, M
%\end{align*}
%\noindent denote the cumulative frequency of term $w_i \in \SET{D}$ over adverts in $\SET{T}$.
%
%Given training data for a sensitive topic category $\SET{T}(c) \subset{\SET{T}}, \ c\in\SET{C}$  -- let:
%\begin{align*}
%c_i &=\sum_{{a}\in {T(c)}} \phi_{\SET{D}}(w_i | {a}), \ i=1, \ldots, M
%\end{align*}
%\noindent denote the cumulative frequency of term $w_i \in \SET{D}$ over adverts in $\SET{T}(c)$.
%
%When a vector of adverts ${a}$ is observed -- let:
%\begin{align*}
%a_i &=\sum_{{a}\in {a}_k} \phi_{\SET{D}}(w_i | {a}), \ i=1, \ldots, M
%\end{align*}
%\noindent denote the cumulative frequency of term $w_i \in \SET{D}$ over adverts in ${a}_k$.
%
%We have, therefore, the the following expression for the estimator $\hat{M}_K(c)$ in \eqref{eqn:estimate:mk:simple}:
%\begin{align}
%\hat{M}_K(c) &= \sum_{j=1}^M \ \left( \frac{c_j \ \cdot \ a_j}{t_j} \right)
%\label{eqn:simple:mkc}
%\end{align}

\begin{table}[tp]
\centering
\caption{Illustrative example estimator values.}\label{tbl:toy:model:2}
\begin{tabular}{@{}L{4cm}@{\hskip 10pt}l@{\hskip 10pt}l@{\hskip 10pt}l@{ }}
\cline{1-4}
$w$ & $\sum_{{a}\in \SET{T}} \phi_{\SET{D}}(w | {a})$ & \multicolumn{2}{c}{$\sum_{{a}\in \SET{T(c)}} \phi_{\SET{D}}(w | {a})$}  \\ 
&  & $c=prostate$ &$\bar{c}=other$  \\  
\cline{1-4}   
prostat, cancer                               & $\frac{5}{12}$                 & $\frac{5}{12}$                 & 0                 \\
%\rowcolor[HTML]{ECF4FF}
%\hline
possibl, learn, here                          & $\frac{1}{6}$                & $\frac{1}{6}$                 & 0          \\
%\rowcolor[HTML]{ECF4FF}
%\hline
treat, suffer                       & $\frac{5}{12}$                 & $\frac{1}{4}$                 & $\frac{1}{6}$            \\ 
%\hline
risk                       & $\frac{5}{12}$                 & $\frac{1}{6}$                  & $\frac{1}{4}$                        \\
%\rowcolor[HTML]{ECF4FF}
%\hline
revers, natur, lifetim                        & $\frac{1}{6}$                & 0                & $\frac{1}{6}$    \\              
\cline{1-4}
%\end{tabularx}
\end{tabular}  
\end{table}
%\normalsize
Consider the following illustrative example.   Let $\SET{C} = \{\text{\c{prostate}}\}$ (\emph{i.e.} we have a single sensitive category), label non-sensitive category $\bar{c}$ as $other$ and suppose the training set (after text pre-processing) is,
\begin{align*}
\SET{T} = \big\{
%\begin{bmatrix*}[l]
&(prostate, \ \{\text{prostat cancer possibl risk learn here}\}), \\
&(prostate, \ \{\text{prostat cancer suffer treat}\}),   \\
&(other, \ \{\text{diabet treatment suffer discov revers natur}\}), \\
&(other, \ \{\text{discov lifetim risk diabet}\})
\big\}
%\end{bmatrix*}
\end{align*}
\noindent Dictionary $\SET{D}$ therefore consists of the terms  \{prostat, cancer, diabet, discov, possibl, learn, here, treat, risk, suffer, revers, natur, lifetim\}.   The \PRI estimator values are given in Table II. %\ref{tbl:toy:model:2}.

An advert with text terms (after filtering)
\begin{align*}
a = \{\text{{patient choos safer treat here}}\}
\end{align*}
is observed.  Since the terms {patient, choos, safer} do not appear in the training data set -- only the terms {treat, here } contribute to $\hat{M}_K(c)$.  We have $\phi_{\SET{D}}(w | {a}) = \frac{1}{5}$ for $w \in \{\text{treat, here} \}$ and so $\hat{M}_K(c)=\frac{8}{25}=0.32$ for $c=$ \emph{prostate}.  For comparison, $\hat{M}_K(c)=\frac{2}{25}=0.08$ for $c=$ \emph{other}. The advert in this example is in fact taken from the Google result page for a probe query during a session where the user is carrying out searches related to prostate cancer.  The high value for $\hat{M}_K(c)$ when $c=$ \emph{prostate} is therefore as expected.

\section{Experimental Setup}
\label{sec:experiment:setup}
\subsection{Hardware/Software Setup}
Data was collected using two Linux virtual machines located in a University domain supporting approximately 9,000 users. 
Custom scripts were written to automate query execution and response collection.  These scripts used Python $v2.7$, BeautifulSoup $v4.3.2$  for HTML processing and phantomjs $v1.9.8$ for browser automation.  
Analysis of results was performed on a $2.3$ GHz Intel Core i7 MacBook Pro. The Python SciKit toolkit $v0.15$ \cite{scikit-learn} was used for text preprocessing.  Numeric processing was performed using the NumPy $v1.8.2$ numerical processing tookit \cite{Idris:2012:NC:2464698}.  
\begin{table}[tp]
\rowcolors{2}{gray!25}{white}
\begin{scriptsize}
\tbl{Categories and associated keyword terms .\label{tbl:scripts}}{%
\begin{tabular}{  l L{0.75\columnwidth} @{}}
\hline
\textbf{Category}        & \textbf{Keywords }        \\ 

\cline{1-2}
\c{anorexia}          & nerves eating disorder body image binge diet weight \oldchange{lose} fat                                                                                                                                                                                                                      
%\rowcolor[HTML]{ECF4FF}
\\ %\cline{1-2}
\c{bankrupt}      & bankrupt insolvent bad credit poor credit clear your debts insolvency payday insolvent any purpose quick cash benefits low income                 
%\rowcolor[HTML]{ECF4FF}
\\ %\cline{1-2}
\c{diabetes}         & diabetes mellitus hyperglycaemia blood sugar insulin resistance                                                                        
%\rowcolor[HTML]{ECF4FF}
\\ %\cline{1-2}
\c{disabled}         & disabled special needs accessibility wheelchair                                                                                      
%\rowcolor[HTML]{ECF4FF}
\\ %\cline{1-2}
\c{divorce}           & divorce separation family law                                                                                                          
\\ %\cline{1-2}
%\rowcolor[HTML]{ECF4FF}
\c{gambling addiction}         & uncontrollable  addiction compulsive dependency problem support counselling advice therapist therapy help treatment therapeutic recovery anonymous  
%\rowcolor[HTML]{ECF4FF}
\\ %\cline{1-2}
\c{gay (homosexuality)}           & gay queer lesbian homosexual bisexual transgender LGBT dyke queen homo                                                                          
%\rowcolor[HTML]{ECF4FF}
\\ %\cline{1-2}
\c{location (london)} & london england uk                                                                                                                       
%\rowcolor[HTML]{ECF4FF}
\\ %\cline{1-2}
\c{payday loan}           & default unsecured debt consolidate advice payday cheap                                                                           
%\rowcolor[HTML]{ECF4FF}
\\ %\cline{1-2}
\c{prostate  cancer}       & prostate cancer PSA male urethra urination                                                                                        
%\rowcolor[HTML]{ECF4FF}
\\ %\cline{1-2}
\c{unemployed}       & job seeker recruit search position cv work employment                                                                                                                                                                             
%\rowcolor[HTML]{ECF4FF}
\\ %\cline{1-2}
\c{other}        & Select the top-$50$ queries on Google Trends as examples of non-sensitive queries, excluding terms appearing in sensitive topics.                                                                                                                                                                        
\\ \cline{1-2}
\end{tabular}  }
\end{scriptsize}
%\caption{Categories and associated keyword terms }%
%\label{tbl:scripts}
\end{table}%
%
%%%%%%%%%%%%%%%%%%%%%%%%%%%%%%%%%%%%%%%%%%%%%%
\subsection{User Session Category Selection and Query Creation}
\label{sec:topical:queries}
We select twelve user interest categories to study, detailed in Table~\ref{tbl:scripts}.  Of the eleven sensitive topics, (i) ten are sensitive categories associated with subjects generally identified as causes of discrimination (medical condition, sexual orientation \emph{etc}) or sensitive personal conditions (gambling addition, financial problems \emph{etc}), see for example \cite{Types2:online} (ii) a further sensitive  topic is  related to London as a specific destination location, providing an obviously interesting yet potentially sensitive topic that a search engine might track, (iii) the last topic is a non-sensitive category labeled \oldchange{\c{other}} which is based on the top-$50$ queries taken from Google Trends \cite{google:trends}, providing the catch-all \oldchange{\c{other}} topic representing topics that are not sensitive. \oldchange{The queries selected from Google Trends for the non-sensitive topic do not contain terms appearing in any of the sensitive topic queries.}

For each category apart from \oldchange{other}, a keyword list is created by extracting associated terms from curated sources including Wikipedia (common terms co-occurring on the category page) and Open Directory Project (pages and sub-topics associated with a category).  These are detailed in Table~\ref{tbl:scripts}.   Candidate search queries are then generated for each category by drawing groups of one or more keywords uniformly at random with replacement from the keyword lists.  These candidate queries are manually augmented with common words (and, of \emph{etc}) to yield queries resembling the English language.  In this way a keyword such as fat, for example, might be transformed into a query ``why am i so fat''. Non-sensical or overly robotic queries are removed by manual inspection. 
For the \oldchange{\c{other}} category, queries are taken from the top-$50$ on Google Trends. 

Finally, to construct sequences of queries for use in user sessions, a predefined probe query is inserted at intervals of $1-5$ queries.  In this way we obtain twelve ``scripts'' of queries, each consisting of between $25-40$ queries including the inserted probe queries.   A user session then consists of a single iteration of a single script run from beginning to end.  

%After each script is executed, the environment is cleaned up and then the next script in the set is run. This process is repeated twice daily -- so each scrript is executed twice a day. The same scripts -- with the same queries -- are re-used each time without change so we have a basis for comparison.

\subsection{Selecting Informative Probe Queries}
\label{sec:probe:queries}
A probe query is required to be sufficiently informative that it reveals adaptation in the user-search engine interaction (Assumption 1), but should not overly disturb the search engines responses to user queries (so as to preserve the utility of the search engine for the user).  
To meet these requirements we propose that a good probe query should possess the following general characteristics:
 \begin{description} \label{list:conditions}   
    \item[\textbf{Ambiguity}] It should be meaningful with respect to the sensitive topic  but allow more than one interpretation, so allowing the search engine to choose from a variety of plausible topics.
    \item[\textbf{Consistency}] It should be consistent with the user's information requirement so as not to disturb search engine learning. The probe should not ``surprise'' the search engine.
\end{description}%
Candidate probe query keywords were identified by running each of the scripts in Table~\ref{tbl:scripts} three times, without probe queries, and collecting the response pages.  
We filtered the text in the response pages by stemming terms and removing stopwords. 
Next term frequency analysis of the filtered terms was performed and the top 10 terms 
identified, see Table~\ref{tbl:top10}. 
\begin{table}[tp]
\begin{scriptsize}
\tbl{Top-10 candidate probe terms with term frequency (TF) of occurrence.\label{tbl:top10}}{%
\begin{tabular}{cllllll}
\cline{2-7}
& 
\multicolumn{2}{c}{\textbf{Google}}                                    & \multicolumn{2}{c}{\textbf{Bing}}                                         & \multicolumn{2}{c}{\textbf{Both}}                                         \\ 
\cline{1-7}
\multicolumn{1}{c}{\textbf{Rank}} & \multicolumn{1}{l}{\textbf{Term}} & \multicolumn{1}{l}{\textbf{TF}} & \multicolumn{1}{l}{\textbf{Term}} & \multicolumn{1}{l}{\textbf{TF}} & \multicolumn{1}{l}{\textbf{Term}} & \multicolumn{1}{l}{\textbf{TF}} \\ 
\cline{1-7}
\rowcolor{gray!25} 
\textbf{1}                          & help                              & 4.37                                   & help                              & 4.62                                   & help                              & 4.49                                   \\
\rowcolor{gray!25} 
\textbf{2}                          & advice                            & 4.32                                   & advice                            & 3.45                                   & advice                            & 4.02                                   \\
\rowcolor{gray!25}
\textbf{3}                          & symptom                           & 1.81                                   & symptom                           & 2.38                                   & symptom                           & 2.04                                   \\
\rowcolor{gray!25}
\textbf{4}                          & cause                             & 0.90                                   & check                             & 0.77                                   & cause                             & 0.82                                   \\
\rowcolor{gray!25}
\textbf{5}                          & homecare                          & 0.60                                   & cause                             & 0.68                                   & person                            & 0.53                                   \\
\textbf{6}                          & offer                             & 0.54                                   & person                            & 0.60                                   & checker                           & 0.49                                   \\
\textbf{7}                          & person                            & 0.48                                   & plan                              & 0.58                                   & check                             & 0.48                                   \\
\textbf{8}                          & answer                            & 0.48                                   & checker                           & 0.58                                   & sign                              & 0.45                                   \\
\textbf{9}                          & gamble                            & 0.44                                   & sign                              & 0.57                                   & offer                             & 0.43                                   \\
\textbf{10}                         & checker                           & 0.43                                   & hiv                               & 0.56                                   & homecare                          & 0.37                         \\
\cline{1-7}      
\end{tabular} }
\end{scriptsize}
%\captionsetup{width=\columnwidth}
%\caption{Top-10 candidate probe terms with term frequency (TF) of occurrence. }
%\label{tbl:top10}%
\end{table}%

It can be seen that the top-$4$ words appearing in both Google and Bing search results are \{help, advice, symptom, cause\} and that these are significantly more frequent than lower ranked terms.  Additionally these terms are in the top-$5$ for each of Google and Bing individually.   We use these keywords to form two probe queries: ``{symptoms and causes}'' for disease and medical topics and ``{help and advice}'' for non-medical topics.

As a rough test of the ambiguity requirement for a probe query discussed in Section~\ref{sec:priv:disclose:search}, we used the number of results indicator provided by each search engine.  We recorded the number of results $N(c)$ returned from querying for sensitive topic $c$ and also the number of results $N(c, p_j), j=1,2$ returned when each of the candidate probe queries is appended to the queries for topic $c$ (with $p_1$=``symptoms and causes'' and $p_2$ =``help and advice'' ).  We expect $N(c, p_j)<N(c)$ since the extra query text will narrow the query to some extent.  However, we would like to avoid this narrowing being too great, \emph{e.g.} we would certainly like to avoid $N(c, p_j)=0$.  The values measured are reported in Table~\ref{tbl:ambiguity} for Google.  Also reported in this table is the ratio $\hat{P}(c | p_j) = \frac{ N(c, p_j) }{ N(c) }$.  It can be seen that $\hat{P}(c | p_1)=0$ for the \c{bankrupt} topic and has a low value for \c{gambling}, \c{gay} and \c{unemployed}.  In contrast, for these topics $\hat{P}(c | p_2)$ has a fairly high value.  This therefore indicates the use of the ``help and advice'' probe query for non-medical topics rather than the ``symptoms and causes'' probe query, which seems intuitive.  Based on Table~\ref{tbl:ambiguity} the ``help and advice'' probe query also seems reasonable for use with medical topics, and $\hat{P}(c | p_1)$ (corresponding to the ``symptoms and causes''  probe) is also reasonable for these topics.   Again, this is as might be expected.

\begin{table}[tp]
\begin{scriptsize}
\tbl{Approximate result numbers returned by Google on different topics and for different choices of probe query. Counts are in units of millions.\label{tbl:ambiguity}}{%
\begin{tabular}{@{}ll@{\hskip 1cm}l@{\hskip 1cm}l@{\hskip 2cm}l@{\hskip 1cm}l@{}}
\cline{1-6}
                  &      & \multicolumn{2}{l}{$p_1=\text{'symptoms and causes'}$} & \multicolumn{2}{l}{$p_2=\text{'help and advice'}$} \\ 
\cline{1-6}
Topic = c         & N(c) & N(c,$p_1$) & $\hat{P}(c|p_1)$ & N(c, $p_2$) & $\hat{P}(c|p_2)$ \\ \cline{1-6}
\c{anorexia}          & 28.5 & 0.834      & 3\%               & 1.78        & 6\%               \\
\c{bankrupt}         & 86.9 & 0.434      & 0\%               & 48.6        & 56\%              \\
\c{diabetes}          & 267  & 66.5       & 25\%              & 114         & 43\%              \\
\c{disabled}          & 506  & 26         & 5\%               & 159         & 31\%              \\
\c{divorce}          & 185  & 11.1       & 6\%               & 79.7        & 43\%              \\
\c{gambling}          & 103  & 0.526      & 1\%               & 30.6        & 30\%              \\
\c{gay}               & 782  & 9.53       & 1\%               & 119         & 15\%              \\
\c{location (london)} & 1930 & 72.2       & 4\%               & 373         & 19\%              \\
\c{payday}            & 70.3 & 45.9       & 65\%              & 6.57        & 9\%               \\
\c{prostate}         & 83.3 & 14.7       & 18\%              & 12.5        & 15\%              \\
\c{unemployed}        & 54.8 & 0.619      & 1\%               & 48.1        & 88\%              \\ \cline{1-6}
\end{tabular} 
}
\end{scriptsize}
%\captionsetup{width=\columnwidth}
%\caption{Approximate result numbers returned by Google on different topics and for different choices of probe query. Counts are in units of millions.}
%\label{tbl:ambiguity}%
\end{table}%

\subsection{User Click Emulation}

To reduce the appearance of robotic interaction, the script automation program inserts a random pause of 1 to 10 seconds between queries, see Table~\ref{tbl:verbatim:location} for an example. After remaining 5 seconds on a clicked link page, the browser ``back'' button is invoked  to navigate back to the search result page. 

To emulate user clicking, we adopt the following user click model.  Given the response page generated in response to a query, for each search result and advert we calculate the Term-Frequency (TF) of the visible text with respect to the keywords associated with session interest category, see Table~\ref{tbl:scripts}.   When the score is $TF > 0.1$, the item is clicked, otherwise it is not clicked.   As mentioned in Section~\ref{sec:priv:disclose:search}, search results in response to probe queries are not clicked. 

%For the random script we concatenated all the keywords from all 11 sensitive topics to create a super-keyword set so we could use the same TF mechanism to test for random chance occurrences of sensitive topics.

%We refer to this click model as the \emph{TF click model} in this paper.

\begin{table}[tp]
\centering
\caption{Example query script. The command \texttt{!wait n} instructs the Python script to wait $n$ seconds}
\label{tbl:verbatim:location}
\begin{multicols}{2}
\begin{verbatim}
! keywords: london england uk  
! probe: help and advice
help and advice
! wait 7
weather forecast for  london
! wait 5
find hotels in london city
! wait 3
help and advice
! wait 7
cheap hotels in london
! wait 10
hotels in regents park cheap
! wait 7
marriott courtyard regents park
! wait 4
help and advice
! wait 7
things to do london next week
! wait 5
regents park hotels
! wait 7
get cheap london show tickets
! wait 7
shows on london now
! wait 5
tickets  london shows
! wait 7
help and advice
\end{verbatim}  
\end{multicols}
\end{table}%

\subsection{Data Collection}
\label{sec:data:collection}
Data was gathered and analysed from the Google Search\footnote{\url{www.google.com}} and Bing\footnote{\url{www. bing.com}} search engines. Data was gathered over a period of 4 weeks during November and December 2014. Scripts were executed daily in the morning and evening over 28 days. 

We took a number of precautions to minimise interactions between runs of each script -- cleaning cookies, history and cache before and after scripts, \oldchange{terminating the session and logging the user out,} and waiting for a minimum of twenty minutes between runs to ensure connections  are reset or timed out.  All scripts were run for $3$ registered users and $1$ anonymous user, and for both the Google and Bing search engines, yielding a data set consisting of $37,134$ queries and response.

The data was partitioned into training and test data sets, see Table~\ref{tbl:training_data}. 
%\textbf{**should we also use different queries for the training data and compare the impact on performance ?}
The test data contains $28$ separate runs of each of the $12$ test scripts.
%
%\oldchange{
%We recognized very early in our experiments that individual query level labeling could give rise to potential keyword bias. Take the example of a probe query, as an extreme case – we use the same probe queries across multiple topics – so should a user label a probe query as `sensitive’ or `non-sensitive’? 
%}
%
\oldchange{For training and performance evaluation we labeled all queries in a session with the intended topic of the session as given by the query script used.  For example, all queries from a session about \c{prostate} are labeled as \c{prostate} or \c{sensitive}, including probe queries.  In this respect the labels capture the intended behavior, rather than attempting an individual interpretation of specific query keywords durring a user session.
}
%In brief our process is the following:
%\begin{description}
%
%\item [Training Phase] PRI is trained using an automatically labeled training set containing adverts labeled as sensitive or non-sensitive based on the intent of the query session.
%
%\item [Pre-test Phase] To compute test accuracy scores, we automatically assign an \emph{expected label} to each test case based on whether the query script used to generate it. 
%
%\item [Test Phase] The PRI tool uses the model constructed from the training data to assign values of $M_{k}(c)$ – one each for sensitive and non-sensitive, or for each topic in the case of multi-topic labeling. 
%
%\item [Prediction Phase] We predict a label for new adverts during testing by taking the label corresponding to the larger of the two M(c) scores. 
%
%\item [Results Phase] We compute how accurately the PRI process predicts labels correctly by comparing with the automatically assigned expected labels.
%
%
%\end{description}

\begin{table}[tp]
\begin{footnotesize}
\centering
\tbl{Summary of training and test data sets. $N_{queries}$ is the number of user search queries and $N_{probes}$ the number of probe queries for which data was collected.\label{tbl:training_data}}{%

\begin{tabular} {lcc@{\hskip 1cm}cc}
\cline{1-5}
& \multicolumn{2}{c@{\hskip 1cm}}{\textbf{Training Data Sets}} & \multicolumn{2}{c}{\textbf{Test Data Sets}}\\
\textbf{Name}  & \textbf{ $N_{queries}$} & \textbf{$N_{probes}$} & \textbf{ $N_{queries}$} & \textbf{$N_{probes}$} \\ 
\cline{1-5}
Bing                    & 1,051    & 367  & 10,970   & 3,795      \\
Google                  & 1,343    & 451  & 14,669   & 4,488     \\
\cline{1-5}
\end{tabular} }
\end{footnotesize}
%\captionsetup{width=0.8\columnwidth}
%\caption{Summary of training and test data sets. $N_{queries}$ is the number of user search queries and $N_{probes}$ the number of probe queries for which data was collected.}
%\label{tbl:training_data}
\end{table}%

%The remaining data was divided into $4$ equal test subsets each comprised of equal quantities of data from Bing and Google. Each subset consisted of approximately $23\%$ of the content from Table~\ref{tbl:data:collection} -- resulting in 
%$16,612$ queries containing $5,439$ probe queries and a total of $112,989$ adverts per subset. Analysis was repeated on each of the $4$ subsets for comparison.
%
\subsection{Feature Selection: Adverts or Links?}
\label{sec:feature:select}
Search result pages contain multiple content types, in particular search links and adverts.   For the collected data sets Table~\ref{tbl:compare:features} summarises the percentage change in the text of search links and adverts for each of the interest categories and for each search engine.   Also shown is $\pm$ the standard error in the mean. It can be seen that link text changes very little, less than $3\%$ for Google and $5\%$ for Bing.  In contrast it can be seen that the advert text is much more dynamic with $12.4\% - 65.5\%$ of the advert text changing for Bing and $17.3\% -  39\%$ for Google. 

This data provides quite strong support for Assumption 2 above, namely that it is the adverts which primarily reveal  personalised learning by the search engine.

\begin{table}[tp]
\begin{scriptsize}
\tbl{Average percentage content change per instance of probe query, grouped by topic and search engine.\label{tbl:compare:features}}{%
\begin{tabular}{l@{ \hskip 25pt}cc@{\hskip 25pt}@{ }cc@{ }}
\cline{2-5}
 \multicolumn{1}{l@{ }}{}  & \multicolumn{2}{c}{\textbf{Bing}} 	 & \multicolumn{2}{c}{\textbf{Google}} \\
\cline{1-5}
\textbf{Topic} 	 & \textbf{Advert} 	 & \textbf{Link}	 & \textbf{Advert} 	 & \textbf{Link} \\
\cline{1-5}
anorexia 	 & 65.4\% $\pm$  7.7\% 	 &  3.6\% $\pm$  0.3\% 	 & 34.8\% $\pm$  1.5\% 	 &  0.9\% $\pm$  0.2\% \\
bankrupt 	 & 15.8\% $\pm$  1.5\% 	 &  5.0\% $\pm$  0.3\% 	 & 39.0\% $\pm$  2.5\% 	 &  2.0\% $\pm$  0.3\% \\
diabetes 	 & 49.4\% $\pm$ 12.5\% 	 &  3.9\% $\pm$  0.3\% 	 & 39.5\% $\pm$  1.7\% 	 &  0.9\% $\pm$  0.2\% \\
disabled 	 & 12.4\% $\pm$  1.0\% 	 &  3.5\% $\pm$  0.2\% 	 & 17.3\% $\pm$  1.7\% 	 &  2.1\% $\pm$  0.3\% \\
divorce 	 & 15.8\% $\pm$  1.7\% 	 &  4.7\% $\pm$  0.4\% 	 & 22.1\% $\pm$  2.5\% 	 &  2.9\% $\pm$  0.5\% \\
gambling 	 & 15.7\% $\pm$  1.3\% 	 &  4.0\% $\pm$  0.2\% 	 & 34.2\% $\pm$  1.7\% 	 &  1.8\% $\pm$  0.3\% \\
gay   	 & 13.8\% $\pm$  1.3\% 	 &  4.0\% $\pm$  0.2\% 	 & 34.3\% $\pm$  1.8\% 	 &  2.4\% $\pm$  0.3\% \\
location 	 & 16.3\% $\pm$  1.5\% 	 &  4.8\% $\pm$  0.3\% 	 & 25.3\% $\pm$  2.1\% 	 &  2.4\% $\pm$  0.4\% \\
payday 	 & 17.4\% $\pm$  1.4\% 	 &  3.9\% $\pm$  0.2\% 	 & 29.7\% $\pm$  1.7\% 	 &  1.4\% $\pm$  0.3\% \\
prostate 	 & 52.6\% $\pm$  6.8\% 	 &  3.7\% $\pm$  0.3\% 	 & 34.6\% $\pm$  1.4\% 	 &  0.9\% $\pm$  0.2\% \\
\oldchange{unemployed}	 & 14.3\% $\pm$  1.2\% 	 &  4.5\% $\pm$  0.3\% 	 & 22.8\% $\pm$  1.8\% 	 &  2.9\% $\pm$  0.5\% \\
other 	 & 17.8\% $\pm$ 27.9\% 	 &  3.7\% $\pm$  0.2\% 	 & 27.5\% $\pm$  1.5\% 	 &  1.4\% $\pm$  0.2\% \\
\cline{1-5}
\end{tabular}  }
\end{scriptsize}
%\captionsetup{width=0.9\columnwidth}
%\caption{Average percentage content change per instance of probe query, grouped by topic and search engine. }
%\label{tbl:compare:features}
\end{table}
%%Table~\ref{tbl:tf:all:features} compares Term Frequency (`TF') of probe query  in adverts and links from the data in Table~\ref{tbl:training_data}, to decide if adverts and links are changing in a way that our TF click model can exploit. 
%%
%%Both adverts and links have measurable TF scores on average. As expected, links score highest since they should correspond closely to the probe query. However, marginal variation  between instances of probe queries indicates that links provide little useful reliable discrimination power. In the case of links, we found occasional exceptions -- for example 'bankrupt' and 'disabled' in Table~\ref{tbl:tf:all:features}. On inspection, we found fluctuations in the TF score of the links section of result pages were typically due to topical items injected by the search engine -- such as 'News' and 'Maps' -- rather than the links themselves. 
%%%
%%
%%By comparison, Table~\ref{tbl:tf:all:features} illustrates that advert content varies consistently across topics -- confirming our modelling choice of adverts as the most discriminating element for probe comparison.

\section{Detecting Disclosure}
\label{sec:revealing}
As already discussed, our approach is to issue a sequence of probe queries at steps $k\in K$ interleaved amongst the user queries, use the \PRI estimator to estimate $\hat{M}_{k}(c)$, $k\in K$ based on the response to each probe query and then look for significant changes in these $\hat{M}_{k}(c)$ values. To determine whether changes are significant, for each topic $c\in\SET{C}$ we use the mean plus/minus three standard deviations to define a confidence interval (the mean and standard deviation are estimated using the training data). \changeb{The choice of three standard deviations is taken after performing verification testing on the training data before testing. Choosing the number of standard deviations to use is a balance -- too small a number of standard deviations generates excessive ``False Negatives'' while too  large a number of standard deviations results in a larger number of ``False Positives''.}

\subsection{Sensitive -- Non-sensitive Detection}\label{sec:sensitive:non}

We begin by evaluating the performance of this approach for detecting whether learning of any sensitive topics has taken place or not during a query session, without trying to specify which sensitive topics are involved. For this we use the catch-all other topic $\bar{c}$.  Namely, when the estimate $\hat{M}_{k}(\bar{c})$ lies outside its confidence interval during a user session we take this as rejecting the hypothesis that no learning of sensitive topics has occurred during that session. We standardise a query session to consist of the first $5$ probe queries in a run for the purposes of analysis.

\begin{figure*}[t!]
\centering
       %\begin{subfigure}[t]{\textwidth}
       \centering
       \subfloat[][Expect \c{gambling}, detect  \c{other}. ] {
               \includegraphics [scale=0.5]{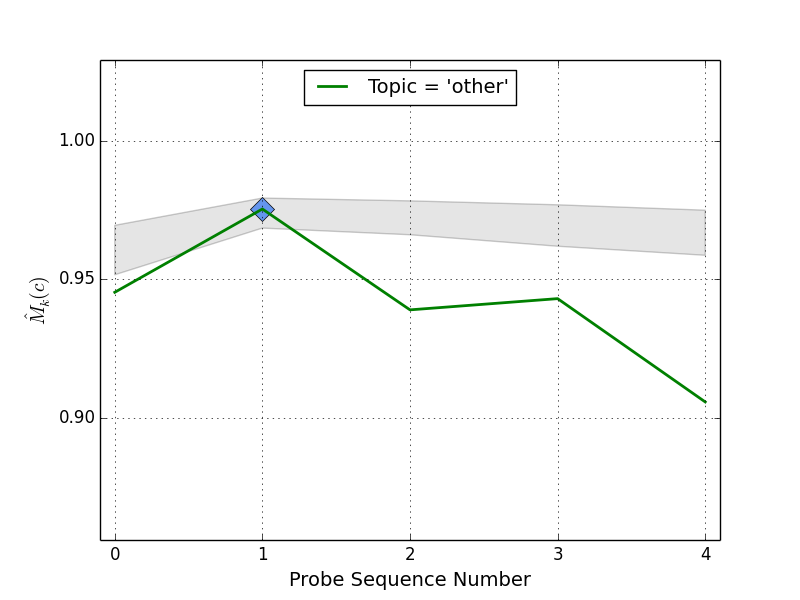} \label{fig:detect:sensitive:google:other}  }
        %\caption{Expect   \replaced[id=PMA]{gambling}{ gambling }  , detect   \replaced[id=PMA]{ other}{ \emph{other} }  }
        %\end{subfigure}% 

       %\begin{subfigure}[t]{\textwidth}
       \subfloat[][Expect \c{gambling}, detect \c{gambling}.  ]{
       \centering
        \includegraphics [scale=0.5]{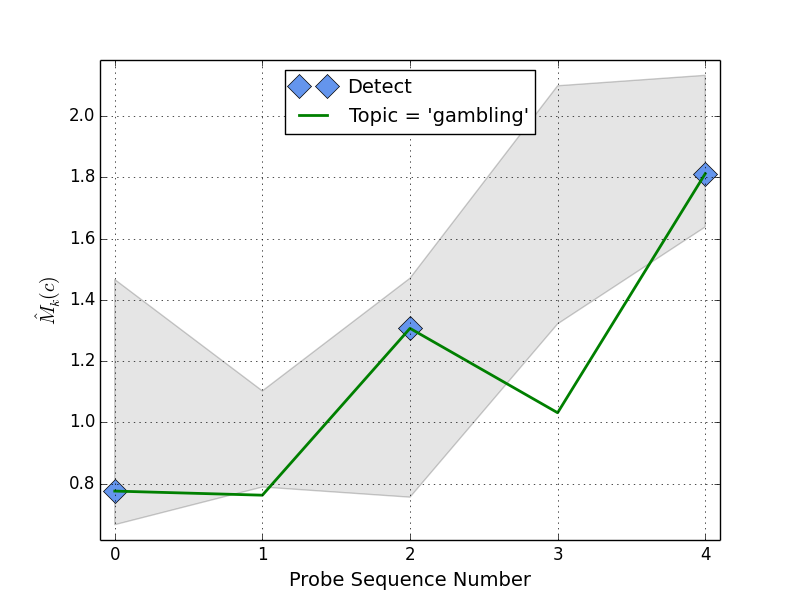}\label{fig:detect:sensitive:google:sensitive} }
        %\caption{Expect   \replaced[id=PMA]{gambling}{ gambling }  , detect   \replaced[id=PMA]{gambling}{ gambling }  }
        %\end{subfigure}%
\caption{Illustrating detection of learning for a user session on topic \c{gambling}. Shaded areas indicate the confidence interval for $\hat{M}_{k}$ for the \oldchange{\c{other}} topic in the upper figure, and for the \c{gambling} topic in the lower figure. Google search engine.}\label{fig:detect:google:sensitive}
\end{figure*}%

\begin{figure*}[t!]
\centering
        %\begin{subfigure}[tp]{\textwidth}
        \subfloat[][Expect \c{gambling}, detect \c{other}.]{
        \centering
        \includegraphics [scale=0.5]{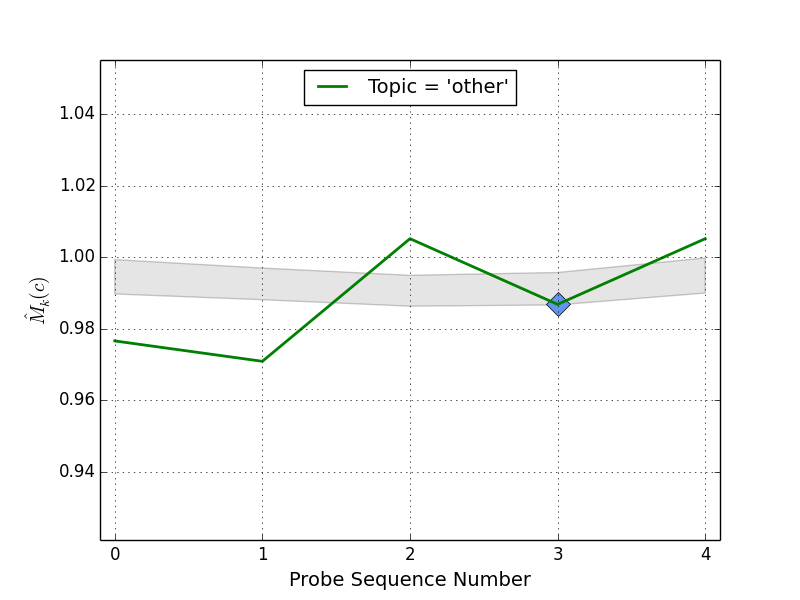}\label{fig:detect:sensitive:bing:other}
        }
        %\caption{Expect   \replaced[id=PMA]{gambling}{ gambling }  , detect   \replaced[id=PMA]{ other}{ \emph{other} }  }

        %\begin{subfigure}[tp]{\textwidth}
        \subfloat[][Expect  \c{gambling}, detect  \c{gambling}.  ]{
        \centering
        \includegraphics [scale=0.5]{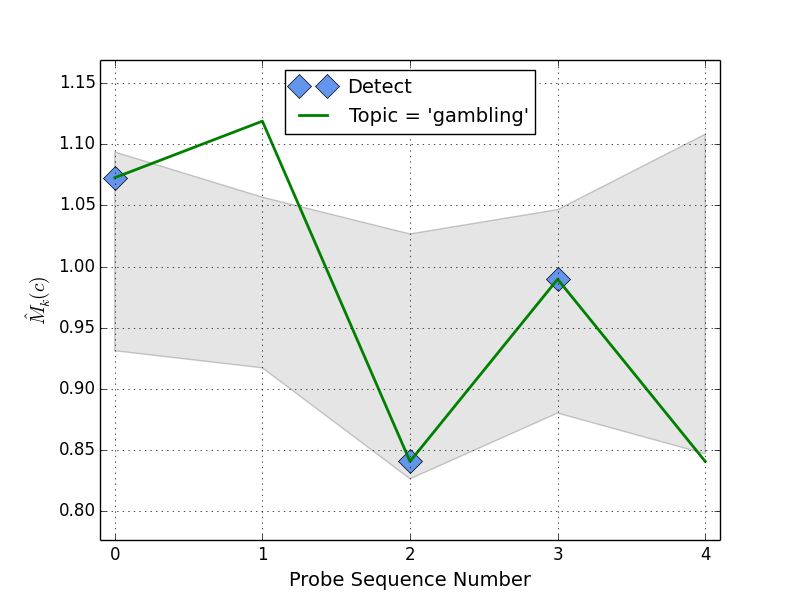}\label{fig:detect:sensitive:bing:sensitive}
        }
        %\caption{Expect   \replaced[id=PMA]{gambling}{ gambling }  , detect   \replaced[id=PMA]{gambling}{ gambling }  }
%\captionsetup{width=0.8\columnwidth}
%\end{center}
\caption{Illustrating detection of learning for a user session on topic \c{gambling}. Shaded areas indicate the confidence interval for $\hat{M}_{k}$ for the \oldchange{\c{other}} topic in the upper figure, and for the \c{gambling} topic in the lower figure. Bing search engine.}
\label{fig:detect:bing:sensitive}
\end{figure*}%

The plots in Figure~\ref{fig:detect:google:sensitive} illustrates this procedure for a user session on the topic   \c{gambling} with the Google search engine.   It can be seen from Figure~\ref{fig:detect:google:sensitive}(a) that   $\hat{M}_{k}$ for the \oldchange{\c{other}}  topic (i.e. $\bar{c}$) quickly leaves its confidence interval as the session progresses (probe $1$ is detected as \oldchange{\c{other}}, however the other probe queries $\{0,2,3,4\}$ lie {outside} the \oldchange{\c{other}} confidence interval).  In comparison, it can be seen from Figure~\ref{fig:detect:google:sensitive}(b) that $\hat{M}_{k}$ for the \c{gambling} topic (i.e. the topic which matches the user session) stays close to the confidence interval throughout the user session.   The corresponding results for the Bing search engine are shown in Figure~\ref{fig:detect:bing:sensitive} and exhibit similar behaviour.

Table~\ref{tbl:google:detect:sensitive} summarises the detection performance on a full set of Bing and Google test data. We declare a positive  detection when at least one probe query in a session of $5$ probes is detected as sensitive.   For user sessions on sensitive topics it can be seen that the detection accuracy is high.  For Google, $100\%$ of user sessions on a sensitive topic reject the hypothesis that no learning of the sensitive topic by the search engine has taken place and so are identified as sensitive.  For Bing the corresponding detection rate is $91\%$.  Recall that this hypothesis testing is being carried out based purely on the adverts in the response pages to user queries, and the queries themselves are not being used.   We manually inspected a sample of the user sessions, confirming the results of Table~\ref{tbl:compare:features}, that the displayed adverts consistently change signficantly over the course of user sessions on sensitive topics. It is therefore reasonable to conclude that learning by the search engine has indeed occurred.  That is, the rejection of the hypothesis that no learning has occured that is reported in Table~\ref{tbl:google:detect:sensitive} appears to be justified.

%\FloatBarrier
\begin{table}[tp]
\scriptsize
\tbl{Measured detection rate of search engine learning of at least one occurence of one or more sensitive topics during a 5 probe session.\label{tbl:google:detect:sensitive}}{%

        \begin{tabular}{@{}llcc@{\hskip 35pt}cc@{}}
        \cline{3-6}
        &             &  \multicolumn{4}{c}{\textbf{Predicted}}  \\ 
        &              & \multicolumn{2}{c}{\textbf{Bing}}     & \multicolumn{2}{c}{\textbf{Google}}    \\
        & & \multicolumn{1}{c}{\textbf{Sens.}}       & \multicolumn{1}{c}{\textbf{Non}\textbf{-sens.}}  & \multicolumn{1}{c}{\textbf{Sens.}}  & \multicolumn{1}{c}{\textbf{Non}\textbf{-sens.}}  \\ 
        \cline{1-6}
        \multicolumn{1}{c}{\multirow{2}{*}{\textbf{Expected}}} & \textbf{Sensitive}    & \multicolumn{1}{c}{91\%}     & \multicolumn{1}{c}{9\%} & \multicolumn{1}{c}{100\% }    & \multicolumn{1}{c}{0\%} \\
        \multicolumn{1}{c}{} & \textbf{Non-sensitive} & \multicolumn{1}{c}{1\%}     & \multicolumn{1}{c}{99\%} & \multicolumn{1}{c}{1\%}   & \multicolumn{1}{c}{99\%}  \\                                                                            
        \cline{1-6}
        \end{tabular} }
\end{table}%

Table~\ref{tbl:google:detect:sensitive} also shows the percentage of user sessions which are sensitive but which are flagged as non-sensitive, which can be interpreted as the false negative rate.  For Google, no sensitive sessions are classed as non-sensitive, and for Bing 9\% are classes as non-sensitive.  Also shown in the table is the percentage of user sessions which are non-sensitive but are flagged as sensitive, which can be interpreted as the false positive rate. This is low at 1\% for both search engines.  A manual inspection of the data shows that the first probe in a session can be misdetected sometimes, demonstrating a \emph{topic lag} effect after there is a change in topic. The influence of the first probe makes it difficult to distinguish sensitive/non-sensitive based on observation of a single step. We will discuss misdetection in detail in Section~\ref{sec:time:learn}.

Overall, the results in Table~\ref{tbl:google:detect:sensitive} indicate that the proposed approach can correctly identify potential privacy concerns for sensitive topics while keeping noise levels from false positive detections low.

We comment briefly on the difference in Table~\ref{tbl:google:detect:sensitive} in the measured False Negative rates for the two search engines.  This difference is at least partially explained by two factors.  The first is that Bing seems to be slower at  adapting to changes in session topic than Google, see Section~\ref{sec:time:learn}.  This apparent difference in adaptation rate is also observable by comparing Figures \ref{fig:detect:google:sensitive}(b) and \ref{fig:detect:bing:sensitive}(b), noting the differences in behaviour of the confidence intervals for the   gambling  topic.  The second factor is differences between the search engines in the range and diversity of the available adverts across the various topics.  For example, analysis of our test data shows that Google has on average $3.3$ unique adverts per probe across all topics whereas Bing has a lower average of $1.7$ unique adverts per probe.  This suggests that Google's dominant position in the search market means it may have a larger advert pool allowing more finely tuned fitting of adverts to detected topics of interest.

\subsection{Individual Sensitive Topic Detection}
\label{sec:probes:detect:changes}
We now evaluate the detection performance for individual sensitive topics.  For each sensitive topic $c$ studied, when (i) the estimated $\hat{M}_{k}(c)$ lies inside the confidence interval for that topic and (ii) $\hat{M}_{k}(\bar{c})$ lies outside the confidence interval for the catch-all \oldchange{other} topic (i.e. $\bar{c}$), then we say that we cannot reject the hypothesis that learning of topic $c$ has occurred.

\begin{table*}[tp]
\caption{Measured detection rate of search engine learning of individual sensitive topics.}
\label{tbl:distinguish:sensitive}
\setlength{\tabcolsep}{0.12cm}
\scriptsize

    %\begin{subtable}[t]{\textwidth}
    \subfloat[][Bing]{
        \centering
        \begin{tabular}{@{}lccccccccccc@{}}
        \cline{2-12}
                                                 & \multicolumn{11}{c}{\textbf{Reference Topic}}                                                                              \\ 
        \multicolumn{1}{l}{}                    & \textbf{anorexia}  & \textbf{bankrupt}   & \textbf{diabetes}  & \textbf{disabled} & \textbf{divorce} & \textbf{gambling}  & \textbf{gay}  & \textbf{location} & \textbf{payday} & \textbf{prostate} & \textbf{unemployed} \\ \cline{1-12}
        \multicolumn{1}{l|}{\textbf{True Detect}}      & 100\% & 98\% & 100\% & 99\% & 99\% & 99\% & 98\%  & 99\% & 99\% & 99\% & 99\% \\
        \multicolumn{1}{l|}{\textbf{True Other}}       & 100\% & 91\% & 93\%  & 93\% & 98\% & 95\% & 100\% & 87\% & 92\% & 96\% & 97\%  \\
        \multicolumn{1}{l|}{\textbf{False Detect}}     & 0\%   & 9\%  & 7\%  & 7\%  & 2\%  & 5\%  & 0\%   & 13\% & 8\%  & 4\% & 3\%   \\
        \multicolumn{1}{l|}{\textbf{False Other}}      & 0\%   & 2\%  & 0\%   & 1\%  & 1\%  & 1\%  & 2\%   & 1\%  & 1\%  & 1\%  & 1\%   \\ \cline{1-12}
        \end{tabular}
        %\captionsetup{width=\textwidth}
        \label{tbl:bing:distinguish:sensitive}
        }

    %\end{subtable}%

    %\begin{subtable}[t]{\textwidth} 
    \subfloat[][Google] {     
        \centering 
        \begin{tabular}{@{}lccccccccccc@{}}
        \cline{2-12}
             & \multicolumn{11}{c}{\textbf{Reference Topic}}                                                                    \\
        \multicolumn{1}{l}{}                    & \textbf{anorexia}  & \textbf{bankrupt}   & \textbf{diabetes}  & \textbf{disabled} & \textbf{divorce} & \textbf{gambling}  & \textbf{gay}  & \textbf{location} & \textbf{payday} & \textbf{prostate} & \textbf{unemployed} \\ \cline{1-12}
\multicolumn{1}{l|}{\textbf{True Detect}}  & 100\%    & 100\%    & 96\%    & 100\%    & 100\%   & 100\%    & 100\% & 99\%     & 99\%   & 99\%     & 100\%      \\
 \multicolumn{1}{l|}{\textbf{True Other}}   & 96\%     & 96\%     & 92\%     & 100\%    & 100\%   & 100\%    & 100\% & 100\%    & 100\%  & 100\%    & 100\%      \\
\multicolumn{1}{l|}{\textbf{False Detect}} & 4\%      & 4\%      & 8\%     & 0\%      & 0\%     & 0\%      & 0\%   & 0\%      & 0\%    & 0\%      & 0\%        \\
\multicolumn{1}{l|}{\textbf{False Other}}  & 0\%      & 0\%      & 4\%      & 0\%      & 0\%     & 0\%      & 0\%   & 1\%      & 1\%    & 1\%      & 0\%        \\ \cline{1-12}
        \end{tabular}
        \label{tbl:google:distinguish:sensitive}
}
    %\end{subtable}%
   
%\captionsetup{width=0.8\columnwidth}
%\end{scriptsize}
\end{table*}

Table~\ref{tbl:distinguish:sensitive} summarises the detection performance for the Bing and Google test data for each of the sensitive topics studied. When evidence of learning of sensitive topic $c$ is detected and the user session is on topic $c$ then we label this a ``True Detect'', otherwise we label this a \changeb{``False Detect''}.  Conversely, when no evidence is found of topic $c$ then when the user session is in fact on topic $c$ we label this a ``\oldchange{False Other}'', otherwise we label this a ``True Other''.  Again, recall that the hypothesis testing here is being carried out based purely on the adverts in the response pages to probe queries.

In the Google results in Table~\ref{tbl:distinguish:sensitive}(b), it can be seen that ``True Detect'' and ``True Other'' results range from $96 - 100\%$ across all sensitive topics. ``False Detect'' results, corresponding to false positives, lie in a range of $0 - 8\%$. ``\oldchange{False Other}'' results, corresponding to  false negatives, are in the range $0 - 4\%$.  We note that topics such as \c{bankrupt} and \c{payday} 
tended to share adverts related to financial services, see next section, making these topics harder to distinguish from one another. This data therefore provides strong support for the assertion that detection of individual sensitive topics is indeed feasible with Google.

Table~\ref{tbl:distinguish:sensitive}(a) presents the corresponding results for Bing.   The ``False Detect'' results, corresponding to false positives, tend to be higher than for the Google data.  
We note that the responses for some sensitive topics overlap in terms of advert content and are not readily differentiated in our data for Bing search (as already noted, in our data set we find that Bing displays fewer unique adverts than Google).  Since our test classifies all non-sensitive topics as \oldchange{\c{other}} then sensitive topics that share adverts with \oldchange{\c{other}} may increase the number of false positives.   Overall, the detection rate for individual sensitive topics is notably high (exceeding 98\%) and the false positive rate remains below 10\% except for the \c{location} topic.

\oldchange{We next test whether probe queries can themselves generate significant levels of false positive sensitive topic detections. We constructed a test script consisting of randomly selected queries from Google Trends into which we injected the previously selected probe queries. This randomised script was executed for both Bing and Google and for each of our user configurations. Relevant result items appearing on non-probe queries were clicked. }
\oldchange{In total $1,264$ probe queries were tested for both Bing and Google using the \PRI framework. Tests yielded a $0\%$  sensitive topic detection rate for any sensitive topic in combinations of search engine and users. We conclude that the selected probe queries do not themselves generate a significant amount of false sensitive topic detection.
}

\subsection{Topic Similarity and Topic Confusion}
\label{sec:epsilon:practice}
Intuitively, we expect that some sensitive topics are similar in the sense that similar adverts tend to be associated with each.  For example, the adverts prompted by the \c{bankrupt} topic, which relates to insolvency, might be expected to have some overlap with the \c{payday} topic, which relates to short-term loans.

\begin{figure}[tp]
\centering
        \subfloat[][Google] {
        \centering
        \includegraphics [scale=0.55]{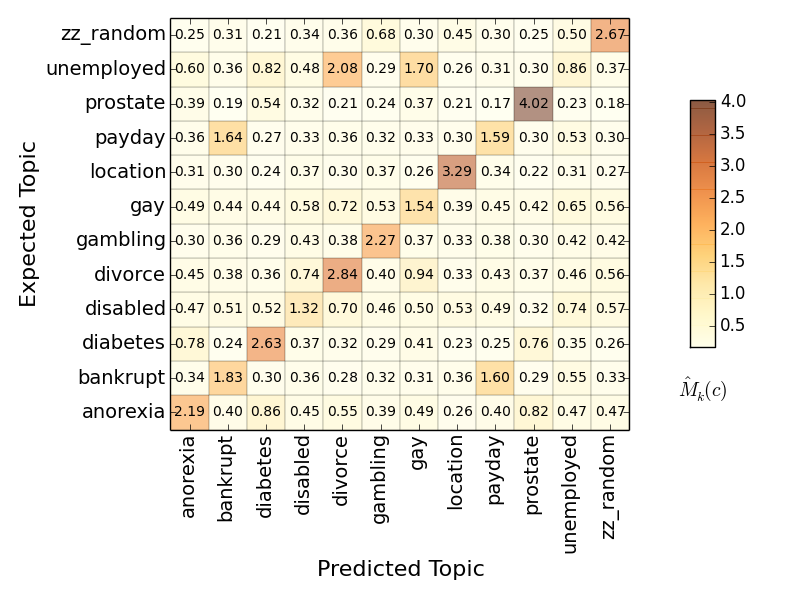}   \label{fig:google_topic_matrix}
        }

       \subfloat[][Bing] {
       \centering
        \includegraphics [scale=0.55]{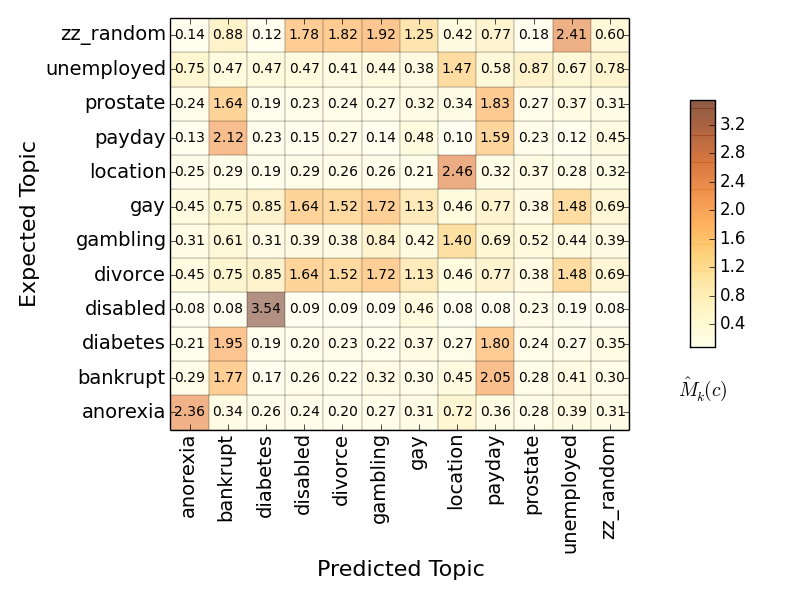}
        \label{fig:bing_topic_matrix}
        }
\caption{Average $\hat{M}_k(c)$ measured by topic.}\label{fig:topic_matrix}
\end{figure}

We can gain some insight into this via the $\hat{M}_k(c)$ estimates for each topic.  Figure~\ref{fig:topic_matrix} shows the average $\hat{M}_k(c)$ measured for each topic $c$ vs the user session topic. That is, \mbox{cell $(i, j)$} shows the average $\hat{M}_k(c)$ measured value attained by \mbox{topic $j$} when running query scripts for reference \mbox{topic $i$}. Each cell is heat-mapped within its row, from brightest for maximum value to darkest for lowest value per row, to improve readability.  Figure~\subref*{fig:google_topic_matrix} shows results for the Google data and Figure~\subref*{fig:bing_topic_matrix} for the Bing data.

For the Google data, it can be seen that the maximum element in each row and column is the diagonal element, as expected from the results presented in the previous section. However, it can also be seen that the \c{payday} topic has a significantly higher $\hat{M}_k(c)$ value than other topics for user sessions on the \c{bankrupt} topic. Similarly, the \c{bankrupt } topic has a significantly higher $\hat{M}_k(c)$ value for user sessions on the \c{payday} topic. Less pronounced, but still evident, is that all health related topics tend have \oldchange{a} higher $\hat{M}_k(c)$ value whenever the user session is on a health topic. For example, \c{diabetes} and \c{prostate} have elevated $\hat{M}_k(c)$ values for user sessions on \c{anorexia}.

For the Bing data in Figure~\subref*{fig:bing_topic_matrix} it can be seen that the results are more complicated. As with Google, the adverts for the \c{payday} and \c{bankrupt} topics show correlated behaviour. Similarly, the adverts for health-related topics tend to be correlated. However, the Bing adverts for the \c{disabled}, \c{divorce}, \c{gambling}, \c{gay} and \c{unemployed} topics also exhibit significant correlation. This is consistent with the results in the previous section where it was observed that topics for Bing appear less readily distinguishable, possibly due to the smaller size of the pool of available adverts.

While the existence of correlation among topics is itself unsurprising, the fact that the proposed approach for detecting search engine learning is able to uncover this correlation provides additional support for the effectiveness of the approach. It also suggests that the potential exists to use the approach to infer additional information from displayed adverts. We explore this further in the following sections.
\subsection{You click -- therefore -- I learn!}
\label{sec:you:click}
In addition to entering queries, users provide feedback to the search engine via the links that they click. Since clicking is an active step, we might expect it to influence search engine learning.  Separate sets of non-click data were collected by running a single iteration of all of the test scripts on both search engines with user clicking turned off.  Table~\ref{tbl:noclick:discussion} shows the percentage change in the average $\hat{M}_k(c)$ score for each test topic with and without user clicking of relevant search results.  It can be seen that all topics had higher $\hat{M}_k(c)$ values when the user clicks on relevant links, suggesting that user clicks are actively used by the search engine for learning.
\begin{table}[tp] 
\begin{scriptsize}
\tbl{Percentage increase in $\hat{M}_k(c)$ by topic for click versus non-click. Google search  data.\label{tbl:noclick:discussion}}{%
\begin{tabular}{lr@{\hskip 25pt}lr@{\hskip 25pt}lr}
\cline{1-6}
\multicolumn{6}{c}{\textbf{Topic -- \% Increase in $\hat{M}_k(c)$}} \\
\cline{1-6}
%\textbf{Topic} & \textbf{\% Increase} & \textbf{Topic} & \textbf{\% Increase} & \textbf{Topic} & \textbf{\% Increase} \\
%\cline{1-6}
\c{anorexia} & 49\%  & \c{divorce} & 153\% & \c{payday} & 62\% \\
\c{bankrupt} & 30\%  & \c{gambling} & 108\% & \c{prostate} & 451\% \\
\c{diabetes} & 417\% & \c{gay} & 158\% & \c{unemployed} & 62\% \\
\c{disabled} & 57\%  & \c{location} & 63\%  & \c{other} & 233\% \\
\cline{1-6}
\end{tabular}  }
\end{scriptsize}
%\captionsetup{width=0.9\columnwidth}
%\caption{Percentage increase in $\hat{M}_k(c)$ by topic for click versus non-click. Google search  data.}%
%\label{tbl:noclick:discussion}
\end{table}%
\subsection{Time to Learn?}
\label{sec:time:learn}
%\begin{figure}[h]
%\includegraphics [width=\columnwidth]{FP_Errors}
%\captionsetup{width=0.9\columnwidth}
%\captionof{figure}{Occurence of classification errors at topic transition boundaries; $N=100$ probe queries -- from both Google and Bing. \textbf{**need to explain this figure, I don't understand what it shows just now}}
%\label{fig:fp:errors}
%\end{figure}%
%

%The results in Table~\ref{tbl:distinguish:sensitive} for logged-in users and in Table~\ref{tbl:distinguish:anonymous} for anonymous users, indicate a search engine does not always identify the correct reference topic, resulting in False Detects and Non-detects. 
Inspection of the test data reveals that correct topic identification sometimes lags by one to two probes at the start of a new user session. This accounts for approximately $70\%$ of cases where ``False Detects'' and ``\oldchange{False Other}'' results are encounted in testing. 
Examination of these cases provides insight into the observed \emph{speed} of recommender learning, and the potential consequences for noise based privacy defences. 
Letting $X$ denote the random variable counting the number of consecutive misclassifications occurring together, then dividing by the total number of misclassifications we can estimate the probability that $X=1$, $X=2$, \emph{etc}.   This data is shown in the first column of Table~\ref{tbl:time:2:learn}.   It can be seen that there are no runs of more than two misclassifications and the average length of a run of misclassifications is,
\begin{align*}
 \E[X; \text{ Bing }] &= 1.77 \\
 \E[X; \text{ Google }] &= 1.05 
 \end{align*} 
 
Letting $Y$ be a random variable indicating the probe sequence number where a ``False Detects'' or ``\oldchange{False Other}''  event \emph{first} occurs, Table~\ref{tbl:time:2:learn} reports the estinated probability that $Y=1$, $Y=2$, \emph{etc}.  As expected the overwhelming majority for ``False Detects'' and ``\oldchange{False Other}'' events happen on the first probe in a session, with $\Pr(Y=1) > 0.90$ for both Bing and Google.
 
The data in Table~\ref{tbl:time:2:learn} therefore suggests that Google search takes an average of $1.05$ probe queries and Bing takes an average of $1.77$ probe queries to re-callibrate learning after a topic change.  On average probe queries in the test data were issued after $4$ user queries.  Hence, Google appears to adapt to a new topic in approximately $4$ queries, while Bing requires approximately $7$ queries.  \oldchange{Rapid recalibration can also be seen in Table~\ref{tbl:fp:by:probe} by looking at sensitive topic  classification recall for Google when successive probe queries are excluded from the calculation. When every probe query is included true positive recall is $62\%$. True positive accuracy improves once the first probe query is excluded and stabilises at $66\%$ thereafter. The false positive rates are low in all cases, falling to $0\%$ when the first three probes are excluded.}

\begin{table}[t]
\tbl{Estimated probabilities of misclassification of various lengths and probe number of first misclassification in a session.\label{tbl:time:2:learn}}{%
\centering 
\begin{tabular}{@{}lc@{}c@{ }@{}lc@{}c@{}}
\cline{1-6}
\multicolumn{3}{l}{\lsplitcell{\textbf{Number of Consecutive Misclassifications (X)}}}
& \multicolumn{3}{l}{\lsplitcell{\textbf{Probe ID of First Misclassification (Y)}}} \\
                              & \textbf{Bing}                   & \textbf{Google}                   &                                 & \textbf{Bing}                     & \textbf{Google}                     \\ \cline{1-6}
$\Pr(X=1)$                    & 0.23                            & 0.95          & $\Pr(Y=1)$                      & 0.92                              & 0.98  \\
$\Pr(X=2)$                    & 0.77                            & 0.05          & $\Pr(Y=2)$                      & 0.03                              & 0.01  \\
$\Pr(X=3)$                    & 0.00                               & 0.00       & $\Pr(Y=3)$                      & 0.04                              & 0.01  \\ 
$\Pr(X=4)$                    & 0.00                               & 0.00       & $\Pr(Y=4)$                      & 0.00                              & 0.00  \\
$\Pr(X=5)$                    & 0.00                               & 0.00       & $\Pr(Y=5)$                      & 0.00                              & 0.00  \\
                     \cline{1-6}
\end{tabular} }
\end{table}%

\begin{table}[t]

\tbl{Recall rate by probe query excluding successive probe queries -- Google.\label{tbl:fp:by:probe}}{%
\begin{tabular}{L{2.5cm}C{1cm}C{1cm}C{1cm}C{1cm}}
\cline{1-5}
                              & \textbf{Include All}                   & \textbf{Exclude $k=1$}                   & \textbf{Exclude $k=1,2$}                                & \textbf{Exclude $k=1,2,3$}                                         \\ \cline{1-5}
\textbf{True Positive}                   & 62\%   & 66\%   & 66\%   & 66\%  \\
\textbf{False Positive}                  & 1\%    & 1\%    & 1\%    & 0\%  \\
                     \cline{1-5}
\end{tabular} }
\end{table}%
%/097_PRI_Probe.py -t paper2/google/training_diverse/ -s paper2/google/SMART/true_positive_click_relevant/

This means that a privacy defence based on random topic changes achieved, for example, by injecting spurious queries, could prove to be ineffective unless the \oldchange{spurious} queries are repeated at intervals of less than every $4$ real queries for Google and $7$ for Bing.  This is a considerable overhead.

\subsection{Logged-in vs Anonymous}
\label{sec:anonymous}
\begin{figure}[tp]
    \centering
        \includegraphics [scale=0.55]{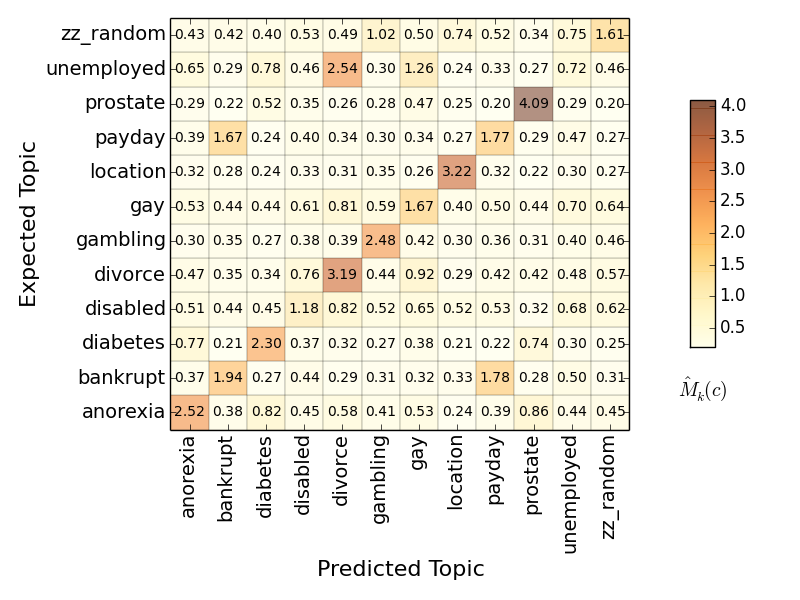}
        %\captionsetup{width=0.9\columnwidth}
        \caption{Average $\hat{M}_k(c)$ by topic. Anonymous user, Google test data}
        \label{fig:anon:google:tm}
\end{figure}%

We collected data for user sessions both when the user is logged-in and when the user is anonymous. As already noted, we clean local caches and user session data between each user session.  

\begin{table}[tp]
\begin{scriptsize}
\tbl{Measured detection rate of search engine learning for an anonymous user.
\label{tbl:google:anon:versus}}{
        \begin{tabular}{llc@{ }cc@{ }c}
        \cline{3-6}
        &             &  \multicolumn{4}{c}{\textbf{Predicted}}  \\ 
        &              & \multicolumn{2}{c}{\textbf{Bing}}     & \multicolumn{2}{c}{\textbf{Google}}    \\
        & & \multicolumn{1}{c}{\textbf{Sensitive}}       & \multicolumn{1}{c}{\textbf{Non-sensitive}}  & \multicolumn{1}{c}{\textbf{Sensitive}}  & \multicolumn{1}{c}{\textbf{Non-sensitive}}  \\ 
        \cline{1-6}
        \multicolumn{1}{c}{\multirow{2}{*}{\textbf{Expected}}} 
                               & \textbf{Sensitive}    & \multicolumn{1}{c}{83\%}     & \multicolumn{1}{c}{0\%} & \multicolumn{1}{c}{100\% }    & \multicolumn{1}{c}{0\%} \\
        \multicolumn{1}{c}{} & \textbf{Non-sensitive} & \multicolumn{1}{c}{17\%}     & \multicolumn{1}{c}{100\%} & \multicolumn{1}{c}{0\%}   & \multicolumn{1}{c}{100\%}  \\                                                                            
        \cline{1-6}
        \end{tabular}}  
\end{scriptsize}
\end{table}

Figure~\ref{fig:anon:google:tm} shows the average $\hat{M}_k(c)$ measured for each topic for the Google search engine when the user is not logged in.  It can be seen that this shows a similar overall pattern to Figure~\ref{fig:topic_matrix}(a), suggesting the search engine is successful at identifying sensitive topics even in the case of an anonymous user. 

\begin{table*}[t]
\centering
\caption{Measured detection rate of search engine learning of individual sensitive topics for an anonymous user.\label{tbl:distinguish:anon}}
\setlength{\tabcolsep}{0.12cm}
\begin{scriptsize}
    \subfloat[Bing] {
        \centering
        \begin{tabular}{@{}lccccccccccc@{}}
        \cline{2-12}
           & \multicolumn{11}{c}{\textbf{Reference Topic}}                                                                              \\ 
\multicolumn{1}{l}{}   & \textbf{anorexia}  & \textbf{bankrupt}   & \textbf{diabetes}  & \textbf{disabled} & \textbf{divorce} &  \textbf{gambling} & \textbf{gay}   & \textbf{location} & \textbf{payday} & \textbf{prostate} & \textbf{unemployed} \\ \cline{1-12}
      \multicolumn{1}{l}{\textbf{True Detect}}      & 100\% & 95\% & 100\% & 98\% & 100\% & 100\% & 96\%  & 100\% & 100\% & 98\%  & 100\% \\
        \multicolumn{1}{l}{\textbf{True Other}}       & 100\% & 83\% & 86\%  & 86\% & 100\% & 100\% & 100\% & 75\%  & 100\% & 100\% & 100\% \\
        \multicolumn{1}{l}{\textbf{False Detect}}     & 0\%   & 17\% & 14\%  & 14\% & 0\%   & 0\%   & 0\%   & 25\%  & 0\%   & 0\%   & 0\%   \\
        \multicolumn{1}{l}{\textbf{False Other}}     & 0\%   & 5\%  & 0\%   & 2\%  & 0\%   & 0\%   & 4\%   & 1\%   & 0\%   & 2\%   & 0\%    \\ \cline{1-12}        
        \end{tabular}        
        %\label{tbl:bing:distinguish:anonymous} 
        }\\
     \subfloat[Google] {
       \centering 
        \begin{tabular}{@{}lccccccccccc@{}}
        \cline{2-12}
          & \multicolumn{11}{c}{\textbf{Reference Topic}}                                                                              \\
\multicolumn{1}{l}{}   & \textbf{anorexia}  & \textbf{bankrupt}   & \textbf{diabetes}  & \textbf{disabled} & \textbf{divorce} &  \textbf{gambling} & \textbf{gay}   & \textbf{location} & \textbf{payday} & \textbf{prostate} & \textbf{unemployed} \\ \cline{1-12}
\multicolumn{1}{l}{\textbf{True Detect}}  & 97\%  & 100\% & 100\% & 100\% & 100\% & 100\% & 100\% & 100\% & 100\% & 100\% & 100\% \\
 \multicolumn{1}{l}{\textbf{True Other}}   & 100\% & 100\% & 92\%  & 100\% & 100\% & 100\% & 100\% & 100\% & 100\% & 100\% & 100\% \\
\multicolumn{1}{l}{\textbf{False Detect}} & 4\%   & 0\%   & 8\%   & 0\%   & 0\%   & 0\%   & 0\%   & 0\%   & 0\%   & 0\%   & 0\%   \\
\multicolumn{1}{l}{\textbf{False Other} } & 3\%   & 0\%   & 0\%   & 0\%   & 0\%   & 0\%   & 0\%   & 0\%   & 0\%   & 0\%   & 0\%       \\ \cline{1-12}
        \end{tabular}
        %\label{tbl:google:distinguish:anonymous}
        }               
\end{scriptsize}
\end{table*}

% Had to hard-code the ref to Table XIII --- \ref{tbl:google:distinguish:anonymous}
Table~\ref{tbl:google:anon:versus} shows the corresponding measured rates for sensitive/non-sensitive topic detection, which can be compared to Table~\ref{tbl:google:detect:sensitive}.   \oldchange{Table XIV}  shows the detection rate for individual topics, which can be compared to Table~\ref{tbl:distinguish:sensitive}.  It can be seen that the detection rates are similar to the results presented previously for logged-in users.  In particular the True Detection rate for individual topics is high e.g. $97 - 100\%$ for Google. 

We conclude that anonymity seems to provide little protection within an individual query session.  \oldchange{The results of Section~\ref{sec:time:learn} show} that the users previous search history is not really required to infer the topic of a sessions, the session itself is enough.

\section{Conclusions and Discussion}
%%%%%%%%%%%%%%%%%%%%%%%%%%%%%%%%%%%%%%%%%%%%%

With \oldchange{\mbox{$\epsilon$-indistinguishability}} as a practical model for detection of user privacy risk, we show that this is readily implementable with available open tools that are simple to apply and provide highly accurate results. An appealing aspect is the use of openly available resources -- Bing and Google search -- a feature often missing in traditional privacy research where concerns over data disclosure limit access to potentially sensitive test data sources. 

\changeb{ 
The results in this paper suggest a number of interesting avenues for future research into how to construct effective counter-measures to sensitive user profiling. The observations of learning lag in Section~\ref{sec:time:learn}, the observation of low, but non-zero false positives in experimental results in Section~\ref{sec:revealing},  and the effect of clicking on learning in Section~\ref{sec:you:click} appear promising for future research into effective counter-measures.}

\oldchange{
Our current experiments focus on browser-based interaction from a personal computer. In the interests of simplicity, we excluded platforms, such as mobile devices, from our current investigation. Mobile devices represent an interesting area for further investigation. The physical size of the screen on mobile and tablet devices increases the urgency to target adverts, while access to finer-grained location and usage data provides even more opportunity to target recommendations. 
}

\changeb{
The user-browser interaction model chosen for this paper in Section~\ref{sec:topical:queries} is based on clicking on links and subsequently navigating back to the original search page by pressing the ``back'' button. The user-browser interaction model can be extended in interesting ways: by allowing the user to spawn new browser tabs and new windows for example. We also removed cookies stored by the search engine during sessions to ensure observations were related to individual sessions. Allowing cookies to persist across sessions, and so potentially preserve learning effects across sessions, is another interesting variation of user-browser interaction that merits investigation. 
}

\changeb{
In Section~\ref{sec:topical:queries} the method for selecting probe queries in this paper based on high-occurrence terms is discussed. The resulting choice of probe queries in Section~\ref{sec:topical:queries} is dependent on the choice of topics. A different choice of topics may necessitate a different choice of probe query. The approach taken in Section~\ref{sec:topical:queries} is to select keywords with the highest term frequency across result pages. It is possible that other combinations of keywords may generate more informative probes, allowing more sensitive detection of search engine adaptation. How to select the most effective and informative probe queries, aligned with an individual user's choice of topics, in a way that is not overly onerous for a user is a subject for future research. 
}

\changeb{
In this paper, training set data was derived by selecting subsets of test data and deriving a dictionary of terms $\SET{D}$. In a practical implementation, a predefined dictionary containing words appearing in adverts can be substituted for $\SET{D}$. Frequency data for terms can then be learned either in batch mode from a pre-labelled set of example adverts, or in online mode by user labelling with new terms being added to the dictionary of terms as they are encountered. In a practical implementation some degree of online learning is desirable as adverts change over time. For example, comparison of data gathered over six month intervals indicates that as much as $30\%$ of terms used in adverts may change over that period of time.
} 

\changeb{
In this paper, our goal is to inform the user by \emph{detecting} evidence of privacy disclosure. In this way we hope to raise individual awareness of privacy concerns from online personalisation. A natural next step for future research is to ask what actions an individual can take to assert control in the face of privacy concerns? The varying and contextual nature of individual privacy concerns, explored in \cite{boyd2011talk}, \cite{Panjwani:2013:UPD:2470654.2466470} and \cite{Agarwal:2013:ERU:2501604.2501612}, suggests that this is a challenge for future research.  
}

We view this paper as a starting point towards practical user privacy in the face of ever-evolving and more powerful online  systems. Future avenues of research include:  looking beyond search engines to other recommender systems where content types other than adverts may provide better content for adaptation detection in the case of other recommender systems; extending \oldchange{$\epsilon$-indistinguishability} to incorporate more complex user interaction models; constructing  effective user privacy defences by exploiting observations of topic similarity and confusion encountered in our experiments, and, investigating how \textbf{PRI} performs for different models of contextual advert selection such as semantic or sense-based techniques that employ non-keyword based selection techniques to select adverts.
In conclusion, our results indicate that evidence of adaptation is easy to find.  Indeed it is mandated to maximise shareholder value.  This suggests that there is an ``Elephant in the Room'' for privacy in the face of sophisticated of modern commercial internet systems.  Namely, focusing on personal de-identification is to risk missing the larger threat of distinguishability. Our observation that such sensitive topic profiling persists even for anonymous users helps to further underline the nature of the privacy threat.

\endgroup
%%%%%%%%%%%%%%%%%%%%%%%%%%%%%%%%%%%%%%%%%%%%%
\bibliographystyle{ACM-Reference-Format-Journals}
\bibliography{general.Accepted}

\end{document}